\documentclass[twocolumn,twocolappendix]{aastex631}

\usepackage{svg}
\usepackage{subfigure,graphicx}
\usepackage{amsmath}
\usepackage{graphbox}
\usepackage{comment}
\usepackage{threeparttable}
\usepackage{color}

\definecolor{purple}{rgb}{0.9, 0.0, 0.9}

\definecolor{green}{rgb}{0, 0.6, 0.1}

\newcommand\ionf[2]{[#1$\;${\scshape{#2}}]}
\usepackage{CJK}

\graphicspath{{./figs}}
\DeclareGraphicsExtensions{.pdf}

\begin{document}

\title{The physical origin of positive metallicity radial gradients in high-redshift galaxies: insights from the FIRE-2 cosmological hydrodynamic simulations}

\correspondingauthor{Xin Wang}
\email{xwang@ucas.ac.cn}

\author[0009-0005-8170-5153]{Xunda Sun}
\affil{School of Astronomy and Space Science, University of Chinese Academy of Sciences (UCAS), Beijing 100049, China}

\author[0000-0002-9373-3865]{Xin Wang}
\affil{School of Astronomy and Space Science, University of Chinese Academy of Sciences (UCAS), Beijing 100049, China}
\affil{Institute for Frontiers in Astronomy and Astrophysics, Beijing Normal University,  Beijing 102206, China}
\affil{National Astronomical Observatories, Chinese Academy of Sciences, Beijing 100101, China}

\author{Xiangcheng Ma}
\affil{Department of Astronomy and Theoretical Astrophysics Center, University of California Berkeley, Berkeley, CA 94720, USA}

\author[0000-0002-3775-0484]{Kai Wang}
\affil{Kavli Institute for Astronomy and Astrophysics, Peking University, Beijing 100871, People's Republic of China}

\author[0000-0003-0603-8942]{Andrew Wetzel}
\affil{Department of Physics and Astronomy, University of California, Davis, CA, USA 95616}

\author[0000-0002-4900-6628]{Claude-Andr\'{e} Faucher-Gigu\`{e}re}
\affiliation{CIERA and Department of Physics and Astronomy, Northwestern University, 1800 Sherman Ave, Evanston, IL 60201, USA}


\author[0000-0003-3729-1684]{Philip F. Hopkins}
\affil{TAPIR, Mailcode 350-17, California Institute of Technology, Pasadena, CA 91125, USA}

\author[0000-0002-1666-7067]{Du\v{s}an Kere\v{s}}
\affil{Department of Physics, Center for Astrophysics and Space Sciences, University of California San Diego, 9500 Gilman Drive, La Jolla, CA 92093, USA}

\author{Russell L. Graf}
\affil{Department of Physics and Astronomy, University of California, Davis, CA, USA 95616}

\author{Andrew Marszewski}
\affiliation{CIERA and Department of Physics and Astronomy, Northwestern University, 1800 Sherman Ave, Evanston, IL 60201, USA}

\author{Jonathan Stern}
\affil{School of Physics \& Astronomy, Tel Aviv University, Tel Aviv 69978, Israel}

\author{Guochao Sun}
\affiliation{CIERA and Department of Physics and Astronomy, Northwestern University, 1800 Sherman Ave, Evanston, IL 60201, USA}

\author{Lei Sun}
\affil{School of Astronomy and Space Science, University of Chinese Academy of Sciences (UCAS), Beijing 100049, China}

\author{Keyer Thyme}
\affil{Department of Astronomy and Astrophysics, University of Chicago, Chicago, IL 60637, USA}

\begin{abstract}
Using the FIRE-2 cosmological zoom-in simulations, we investigate the temporal evolution of gas-phase metallicity radial gradients of Milky Way-mass progenitors in the redshift range of $0.4<z<3$. We pay special attention to the occurrence of positive (i.e. inverted) metallicity gradients --- where metallicity increases with galactocentric radius. This trend, contrary to the more commonly observed negative radial gradients, has been frequently seen in recent spatially resolved grism observations. The rate of occurrence of positive gradients in FIRE-2 is about $\sim7\%$ for $0.4<z<3$ and $\sim13\%$ at higher redshifts ($1.5<z<3$), broadly consistent with observations. Moreover, we investigate the correlations among galaxy metallicity gradient, stellar mass, star formation rate (SFR), and degree of rotational support. 
Metallicity gradients show a strong correlation with both sSFR and the rotational-to-dispersion velocity ratio ($v_c/\sigma$), implying that starbursts and kinematic morphology of galaxies play significant roles in shaping these gradients.
The FIRE-2 simulations indicate that galaxies with high sSFR (${\rm log(sSFR~[yr^{-1}])}\gtrsim-9.2$) and weak rotational support ($v_c/\sigma\lesssim 1$) are more likely --- by $\sim$15\% --- to develop positive metallicity gradients.
This trend is attributed to galaxy-scale gas flows driven by stellar feedback, which effectively redistribute metals within the interstellar medium.
Our results support the important role of stellar feedback in governing the chemo-structural evolution and disk formation of Milky Way-mass galaxies at the cosmic noon epoch.

\end{abstract}


\section{Introduction} \label{sec:intro}

Metallicity is one of the most fundamental properties of galaxies.
It can be quantified in two elemental phases: stellar metallicity \citep[e.g.][]{Gallazzi2005, Kirby2013}, reflecting the time-averaged abundance across the entire galactic star formation history, and gas-phase metallicity \citep[e.g.][]{Tremonti2004, Lee2006}, which indicates the present state of metal enrichment in the interstellar medium (ISM).
These metrics are essential for understanding the evolution of galaxies.
In particular, the galaxy mass-metallicity relation (MZR) presents a critical observed trend of tight correlation of galaxy stellar mass with both gas-phase metallicity and stellar metallicity. 
Metallicity may be influenced by various feedback mechanisms, including gas accretion, supernova explosions, AGN feedback, merge, and stellar winds \citep{King2015, Wangk2023}.
The MZR indicates that more massive galaxies generally possess higher metallicity, a trend attributed to their enhanced ability to retain metal-enriched gas and convert gas into stars \citep{Wang2017, Wang2020, Wang2022, He2024}.
These findings suggest that the interplay between galaxy mass and metal content is important for understanding galaxy formation and evolution, offering insights into the processes that govern the chemical enrichment of galaxies across the universe \citep{2021ApJ...922..189V,Lian2023}.

The metallicity radial gradient is widely used to study the spatial distribution of metals in galaxies.
Analyzing its response to feedback mechanisms helps us reveal the role these processes play in regulating galactic evolution.
Since \cite{Searle1971}, it has been known that galaxies in the present-day universe are more inclined to show negative gas-phase metallicity gradients, implying that gas is more metal-enriched in the inner galaxy \citep[e.g.][]{Zaritsky1994, vanZee1998}.
Typically, galaxies exhibit negative metallicity gradients, characterized by a decrease in metallicity with increasing radial distance from the galactic center.
Since the density of stars declines more steeply than that of gas, there is a gradual decrease of metallicity in the stellar regions as the radius increases \citep{Ho2015, Ma2017}.
While \cite{WangEC2022_Accretion} established a metal enrichment model which simplified the accretion mechanisms of galactic disks.
This model demonstrates that the negative metallicity gradient is actually facilitated by the inflow of cold gas, providing a different perspective.
\cite{Boardman2022} found that galaxies with larger physical extents at a given stellar mass ($>10\rm M_\odot$) tend to have steeper metallicity gradients, which suggests that galactic feedback significantly shapes these gradients over extended timescales, reflecting the prolonged evolutionary history of these galaxies.
\citet{Stinson2010, Brook2012, pilkingtonMetallicityGradientsDisks2012, Gibson2013} analyzed the variation of metallicity gradients under two distinct feedback models, demonstrating that the "enhanced" feedback model (MaGICC) yields flatter gradients compared to the "conservative" feedback model (MUGS).
FOGGIE \citep{Peeples2019} exhibit a progressive flattening of their gradients as they evolve\citep{Acharyya2024arXiv}, similar to MUGS.
TNG50 \citep{Marinacci2018,Naiman2018,Springel2018,Pillepich2018,Nelson2018} and EAGLE \citep{Crain2015,Schaye2015,McAlpine2016} indicates that galaxy metallicity gradients depend on multiple properties, with more massive galaxies exhibiting flatter gradients and lower mass galaxies showing greater fluctuations \citep{Hemler2021,Tissera2022}.
\citep{Garcia2025} conducted a study of galaxies spanning a broad mass and redshift range in the EAGLE, Illustris, IllustrisTNG, and SIMBA simulations, and suggested that the similarly implemented smooth stellar feedback models in these simulations may lack the strength necessary for efficient metal mixing.
\citet{Tapia2025} highlight the complex interplay between internal dynamics and external gas flows in shaping metallicity distributions, and suggest that disk dynamics can play a significant role in the suppression or modification of metallicity gradients.
According to \cite{Bellardini2021, Bellardini2022}, as galaxies evolve, their disks gradually become rotationally supported, which restricts the radial mixing of metals in gas, leading to a gradual steepening of the radial metallicity gradients over time \citep[also see][for the very recent observational evidence from the MSA-3D survey]{Ju2024arXiv}.
Strong and weak feedback models yield markedly different outcomes, and the underlying physical processes further influence the evolution of metallicity gradients.

Although the majority of galaxies possess negative metallicity gradients, a number of them show positive gradients, as first identified by \cite{Cresci2010} at $z\sim3.0$.
This anomalous pattern, contrary to the usual tendency, indicates higher metal concentrations in the outskirts than in the central regions of these galaxies.
Such distributions are consistent with observations of star-forming galaxies at high redshifts, which generally exhibit metal-enriched gas outflows, expected to result in spatial variations of metal distribution across galaxies.
Subsequent studies have quantified the occurrence of positive metallicity gradients. \citet{Montero2016} reported that approximately $10\%$ of galaxies in the CALIFA nearby galaxy sample show positive gradients.
\citet{Carton2018} found this percentage to be around $8\pm3\%$ at $0.1<z<0.8$ from MUSE.
In addition, \citet{Tissera2022} reported that $20-25\%$ of galaxies at $z\leq2$ from the EAGLE simulations show positive metallicity gradients.
\cite{Schonrich2017} suggests that these positive metallicity gradients are likely caused by high rates of central gas loss and re-distribution processes, such as the re-accretion of metal-enriched material combined with inside-out galaxy formation and near-disk galactic fountaining.
FOGGIE \citep{Acharyya2024arXiv} shows that its positive gradients are caused by close interactions and merging systems.
\cite{Wang2019} reports the first sub-kpc resolution measurements of extremely positive metallicity gradients in two dwarf galaxies at redshift $z\sim2$, demonstrating rapid mass assembly and significant impact of stellar nucleosynthesis and gas outflows on chemical distributions.
\cite{Venturi2024} shows that the redistribution of metals in galaxies may be dominated by mergers.
These phenomena suggest that galactic evolution processes may include a variety of complex feedback mechanisms, resulting in these observed positive gradients that differ from more common scenarios.
Understanding these unique phenomena could provide valuable insights into the diversity of galactic structures and the mechanisms driving their formation and growth.

The Feedback In Realistic Environments (FIRE) project\footnote{See also: \url{https://fire.northwestern.edu/}} \citep{Hopkins2014, Hopkins2018, Hopkins2023} comprises a set of cosmological zoom-in simulations that explore how feedback mechanisms influence galaxy formation, gas distribution, chemical evolution, and morphological development.
In this work, we use eight cosmological zoom-in simulations from the Latte suite of Milky Way-mass galaxies \citep[introduced in][]{Wetzel2016}, part of the FIRE-2 project, to study the relations between gas-phase metallicity radial gradients and other galactic properties, and to examine the proportion of inverse metallicity gradients under various conditions to understand their causes.
The simulated galaxies from the FIRE-1 and FIRE-2 samples have been used to explore the properties of the MZR \citep[see e.g.,][]{Ma2016, Porter2022, Marszewski2024, Bassini2024arXiv} and the radial gradients of elemental abundances \citep[see e.g.,][]{Mercado2021, Porter2022, Orr2023}.
In particular, \citet{Ma2017} studied the diversity of the spatially resolved gas-phase metallicities in galaxies at $0.6<z<3$ using the FIRE-1 simulations.
\citet{Bellardini2021, Bellardini2022} investigated the three-dimensional (3D) elemental abundances pattern in both gas and in stars at birth in FIRE-2 simulations, tracing their formation histories at $z<1.5$.
They demonstrated that metallicity radial gradients at early times typically appear flat and can scatter to positive values; additionally, they showed the consistency of the ISM gradients with observational data at $z=0$.
\citet{Graf2024} showed that the trends of stellar gradients with age in these FIRE-2 galaxies today parallel those measured in the Milky Way.
In this work, aiming at a comprehensive exploration of the physical causes of positive metallicity gradients, we analyze the spatially resolved properties of the FIRE-2 Milky Way-mass galaxies in a wide redshift range of $0.4<z<3$, going further back to the cosmic noon epoch.

The organization of this paper is as follows.
In Section~\ref{sec:method}, we introduce the simulation details of FIRE-2 and describe the methods used to measure the gas-phase metallicity gradients of these simulated galaxies.
Our main results are presented in Section~\ref{sec:result}, and our conclusions are provided in Section~\ref{sec:conclusion}.

\section{Methodology} \label{sec:method}

\subsection{Simulations}\label{sec:simu}
\begin{table*}[!htb]
\caption{Simulation details.}
\begin{threeparttable}
\centering
\label{tab:galapro}
    \centering
    \setlength{\tabcolsep}{8pt} 
\begin{tabular*}{0.95\linewidth}{@{}ccccccccc@{}}
\hline
Name & $M_{\rm halo}$ & $m_{\rm baryon}$ & $m_{\rm dm}$ & $\epsilon_{\rm star}$ & $\epsilon_{\rm dm}$ & $\epsilon_{\rm gas, min}$ & Cosmology & Reference \\
  & (M$_\odot$) & (M$_\odot$) & (M$_\odot$) & ({\rm pc}) & ({\rm pc}) &({\rm pc}) &   & \\
\hline
 \texttt{m12z} & $9.25\times10^{11}$ & 4200 & 21,000 & 3.2 & 33 & 0.4 & Z & \cite{Garrison2019a} \\
 \texttt{m12w} & $1.08\times10^{12}$ & 7100 & 39,000 & 4.0 & 40 & 1.0 & P & \cite{Samuel2020} \\
 \texttt{m12r} & $1.10\times10^{12}$ & 7100 & 39,000 & 4.0 & 40 & 1.0 & P & \cite{Samuel2020} \\
 \texttt{m12i} & $1.18\times10^{12}$ & 7100 & 35,000 & 4.0 & 40 & 1.0 & A & \cite{Wetzel2016} \\
 \texttt{m12c} & $1.35\times10^{12}$ & 7100 & 35,000 & 4.0 & 40 & 1.0 & A & \cite{Garrison2019a} \\
 \texttt{m12b} & $1.43\times10^{12}$ & 7100 & 35,000 & 4.0 & 40 & 1.0 & A & \cite{Garrison2019a} \\
 \texttt{m12m} & $1.58\times10^{12}$ & 7100 & 35,000 & 4.0 & 40 & 1.0 & A & \cite{Hopkins2018} \\
 \texttt{m12f} & $1.71\times10^{12}$ & 7100 & 35,000 & 4.0 & 40 & 1.0 & A & \cite{Garrison2017} \\
\hline
\end{tabular*}
\begin{tablenotes}
\item \hspace{-13pt}Parameters describing the initial conditions for our simulations (units are physical):
\item Name: simulation designation.
\item $M_{\rm halo}$: approximate mass of the main halo at $z = 0$.
\item $m_{\rm baryon}$, $m_{\rm dm}$: initial masses of baryonic (gas or star) and dark-matter particles.
\item $\epsilon_{\rm star}$, $\epsilon_{\rm dm}$: force softening (Plummer equivalent) for star and dark-matter particles.
\item $\epsilon_{\rm gas, min}$: minimum adaptive force softening (Plummer equivalent) for gas cells.
\item Cosmology: cosmological parameters used in the simulation, as follows:
        \item \hspace{13pt}A (`AGORA': $\Omega_{\rm m}$ = 0.272, $\Omega_\Lambda$ = 0.728, $\Omega_{\rm b}$ = 0.0455, $h$ = 0.702, $\sigma_8$ = 0.807, $n_{\rm s}$ = 0.961);
        \item \hspace{13pt}P (`Planck': $\Omega_{\rm m}$ = 0.31, $\Omega_\Lambda$ = 0.69, $\Omega_{\rm b}$ = 0.0458, $h$ = 0.68, $\sigma_8$ = 0.82, $n_{\rm s}$ = 0.97);
        \item \hspace{13pt}Z ($\Omega_{\rm m}$ = 0.2821, $\Omega_\Lambda$ = 0.7179, $\Omega_{\rm b}$ = 0.0461, $h$ = 0.697, $\sigma_8$ = 0.817, $n_{\rm s}$ = 0.9646).
\item Reference: where the simulation is first presented.
\end{tablenotes}
\end{threeparttable}
\end{table*}
Our analysis focuses on the Latte suite galaxies from the FIRE-2 cosmological zoom-in simulations, all modeled as Milky Way-mass isolated galaxy progenitors.
The Latte suite comprises eight simulations as listed in Table~\ref{tab:galapro} \citep[see also][]{Wetzel2023, Garrison2017, Garrison2019a, Samuel2020, Wetzel2016, Hopkins2018}. The simulations were conducted using the \textsc{gizmo} code in mesh-less finite-mass (MFM) mode \citep{Hopkins2015}.
These simulated galaxies were evolved down to redshift $z=0$, yielding a main halo mass of $M_{\rm halo} \sim 10^{12}$ M$_{\odot}$ at $z=0$.
We selected snapshots spanning redshifts $z\sim0.44-3$, covering the stellar mass range of $M_{\ast}\sim 10^8$ to $10^{11} {\rm M}_{\odot}$ for central galaxies only.
Force softening (Plummer equivalent) values are $\epsilon_{\rm star} = 3.2-4.0$ pc for stellar and $\epsilon_{\rm dm} = 32-40$ pc for dark-matter particles. The minimum adaptive force softening for gas cells is $\epsilon_{\rm gas, min} = 0.4-1.0$ pc. Initial masses of baryonic (gas or star) and dark-matter particles are $M_{\rm baryon} = 4200-7100{\rm M}\odot$ and $M_{\rm dm} = 21000-39000{\rm M}_\odot$, respectively.

FIRE-2 incorporates radiative cooling and heating over a wide temperature range (10 to $10^{10}$ K), while tracking the abundances of 11 elements (H, He, C, N, O, Ne, Mg, Si, S, Ca, Fe).
The model includes detailed stellar feedback mechanisms, including supernovae (Type II and Ia), stellar winds (from OB and AGB stars), and radiative feedback, implemented to realistically capture the dynamics and chemical evolution of the ISM.
These processes, crucial for driving galactic winds and regulating star formation, are based on stellar evolution models with rates and energies derived from \textsc{starburst99} \citep{Leitherer1999} and assume a \citet{Kroupa2001} initial mass function.
Collectively, these processes enrich the ISM with metals and play a critical role in galaxy evolution by regulating gas cooling, heating, and star formation.
Additionally, FIRE-2 includes an optional subgrid metal diffusion model to account for unresolved mixing between gas elements. This model adopts a Smagorinsky-type turbulent diffusion prescription, with diffusivity determined by local velocity gradients and the resolution scale \citep{Hopkins2017,Hopkins2018}.

FIRE-2 features an explicit model for the turbulent diffusion of gas-phase metals \citep{Su2017, Escala2018, Hopkins2018}, which addresses the sub-grid scale mixing processes critical for the chemical evolution of galaxies.
This mechanism facilitates chemical exchange between particles, enabling a more realistic representation of metal mixing and distribution within the ISM. 
In \cite{Bellardini2021}, a comparison was made with the version with no subgrid metal diffusion and one with a re-simulation increasing the diffusion coefficient by 10 times. 
The results indicate that while the diffusion coefficient for metal mixing influences the azimuthal distribution, it has only a small effect on the vertical and radial gradients, which is not sufficient to alter the interpretations in this paper.

\subsection{Galaxy definitions} \label{subsec:gala_def}

We focus solely on the most massive halo in each galaxy at redshifts $z \sim 0.44-3$. 
The detailed physical properties of these galaxies are presented in Appendix~\ref{app:gala_info}.

The main galaxy is identified using the iterative shrinking sphere method\citep{Fitts2017}, which locates the galaxy with the highest stellar mass by progressively reducing the search radius centered on local mass concentrations.
We define the center of each galaxy as the stellar mass centroid.
The stellar mass, gas mass, and SFR, relevant to our calculations, are confined to within 10 kpc of this center.
Following \citet{Ma2017}, we define a characteristic radius $R_{90}$ as the radius enclosing 90 percent of the total SFR within 10 kpc, with the SFR averaged over a period of 200 Myr, which provides a more stable and physically meaningful measure, particularly for high-redshift galaxies, which often exhibit clumpy and irregular morphologies.

\begin{figure*}[ht!]
  \hspace{-0.5cm}
\subfigure{
 \label{fig:m12b_den}
 \centering
\includegraphics[align=c,width=0.5\linewidth]{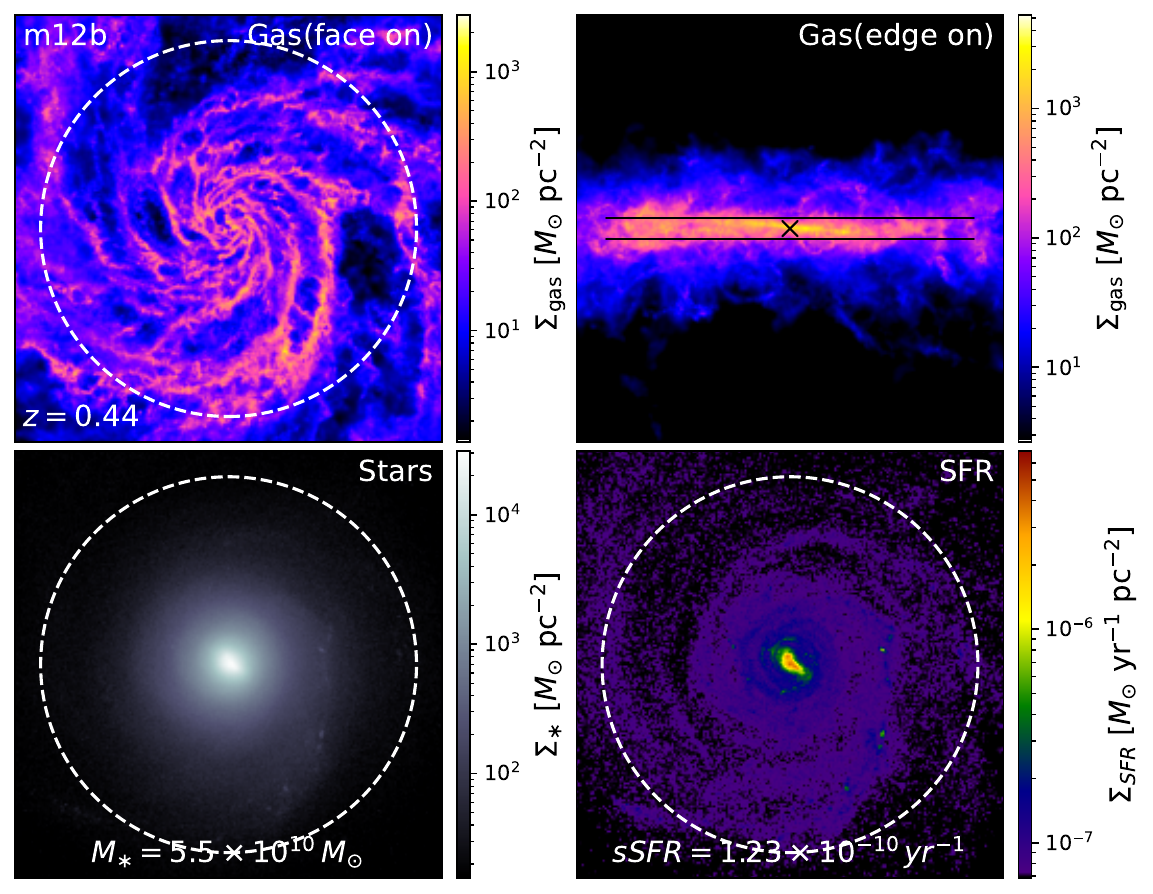}}
  \hspace{0.2cm}
\subfigure{
 \label{fig:m12b_vel}
 \centering
\includegraphics[align=c,width=0.44\linewidth]{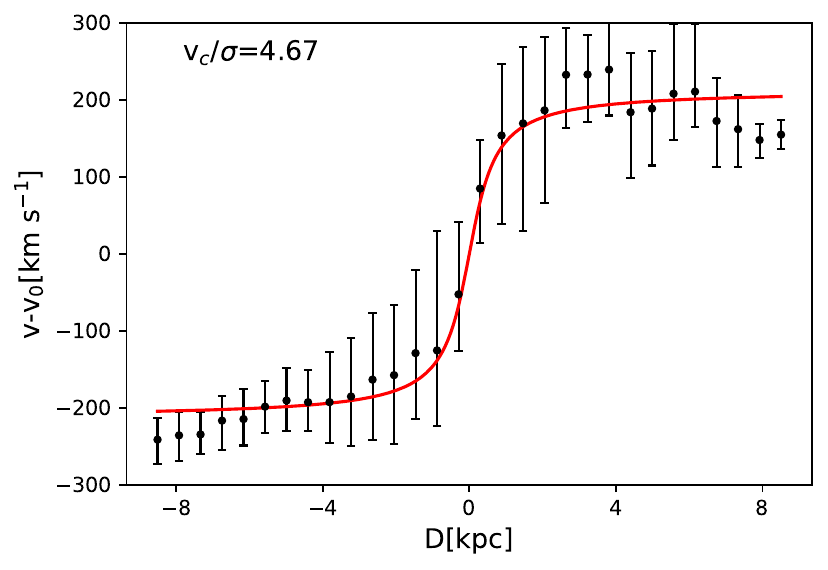}}
 
  \hspace{-0.5cm}
\subfigure{
 \label{fig:m12w_den}
 \centering
 \includegraphics[align=c,width=0.5\linewidth]{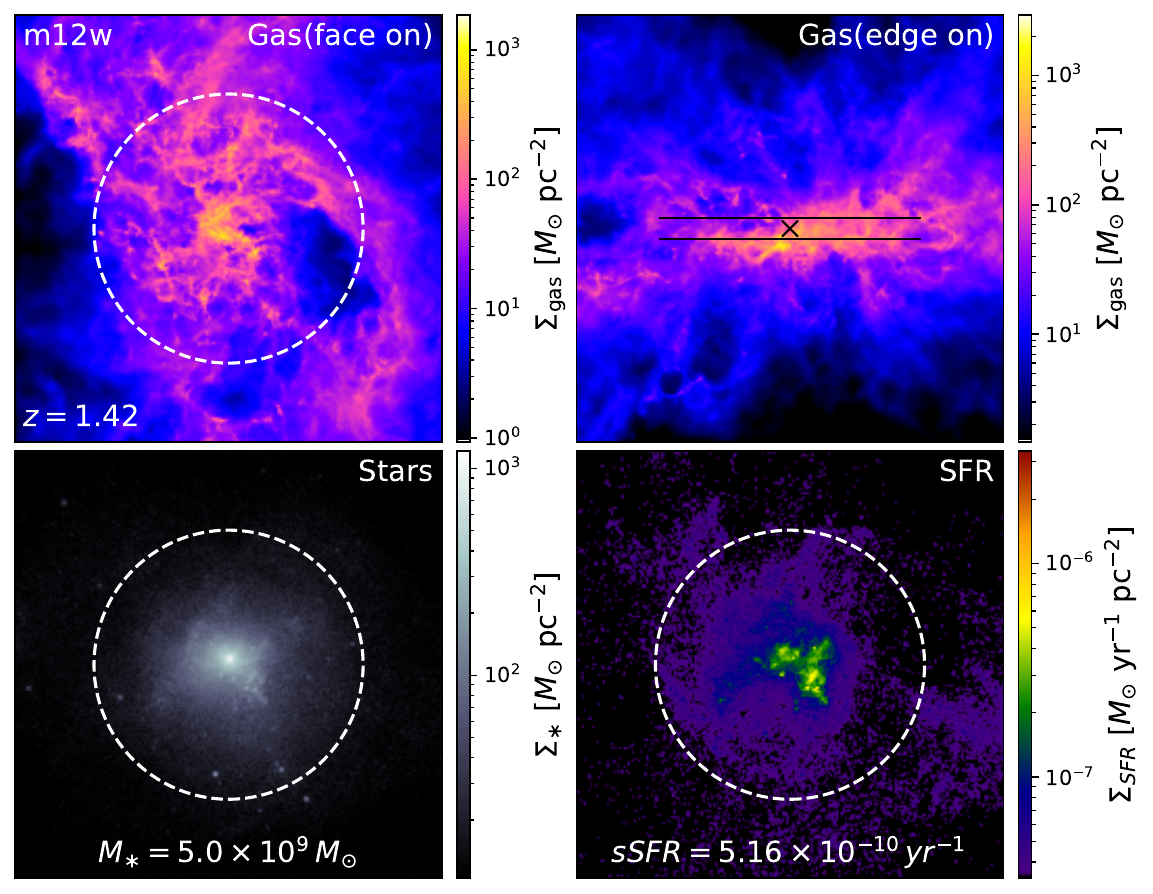}}
  \hspace{0.2cm}
\subfigure{
 \label{fig:m12w_vel}
 \centering
 \includegraphics[align=c,width=0.44\linewidth]{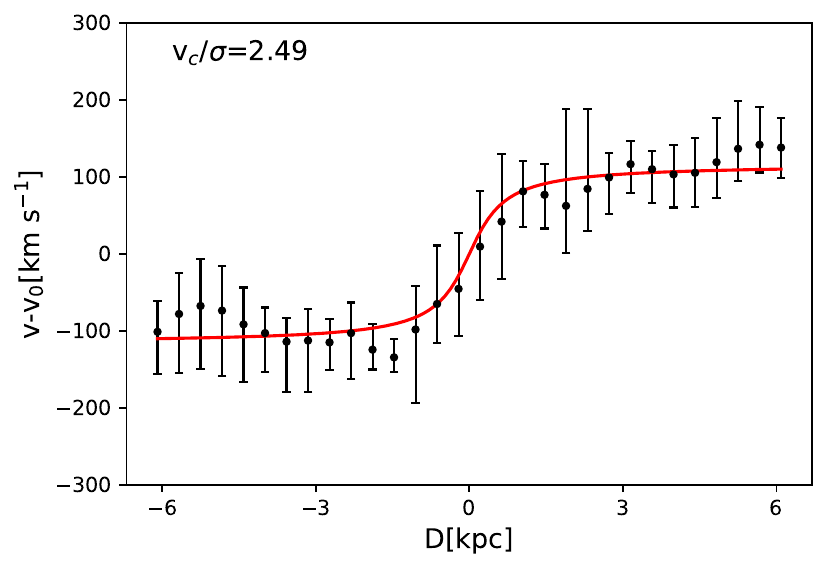}}

  \hspace{-0.5cm}
 \subfigure{
 \label{fig:m12c_den}
 \centering
 \includegraphics[align=c,width=0.5\linewidth]{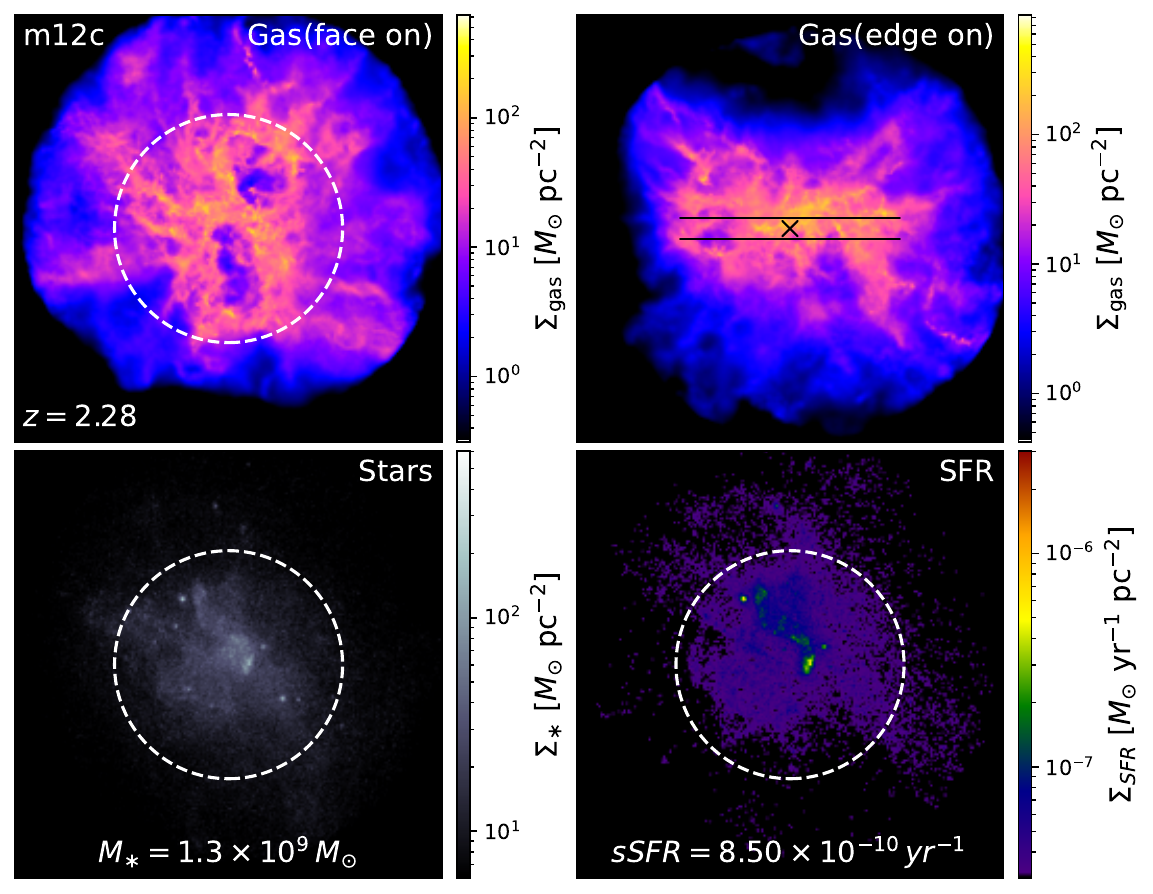}}
  \hspace{0.2cm}
\subfigure{
 \label{fig:m12c_vel}
 \centering
 \includegraphics[align=c,width=0.44\linewidth]{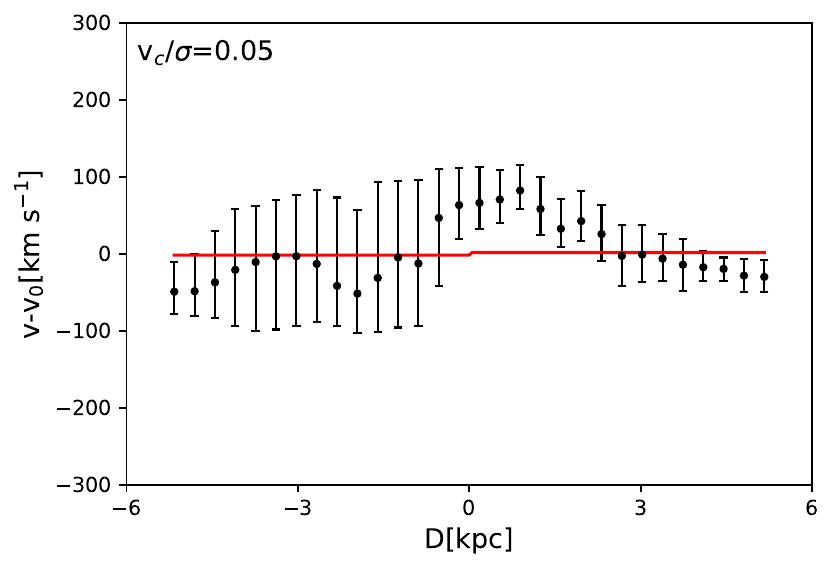}}
\caption{\emph{Left:} Three example galaxies from our FIRE-2 simulations, displaying face-on gas density in the upper left panel, edge-on gas density in the upper right panel, stellar surface density in the lower left panel, and SFR surface density in the lower right panel.
A white circle marks $R_{90}$ defined in Section~\ref{subsec:gala_def}. 
Black lines on the edge-on gas images indicate the long slit from which we extract the gas velocity curves. 
\emph{Right:} Velocity curves of all gas particles extracted from the slit. 
Symbols and error bars represent the velocity and velocity dispersion measurements, respectively, while red lines depict the best-fit results from the arctan function specified in Eq.~\ref{eq:vel}. 
In the upper left, we also display the degree of rotational support ($v_{\rm c}/\sigma$), measured using the 1-$\sigma$ velocity dispersion of the total velocity.
Galaxy \texttt{m12b} exhibits a well-ordered thin disk, \texttt{m12w} has a thick disk, and \texttt{m12c} appears irregular.
\label{fig:gala_def}}
\end{figure*}

In Fig.~\ref{fig:gala_def}, we present three examples of our simulated galaxy sample: \texttt{m12b} at $z \approx 0.44$ (top), \texttt{m12w} at $z \approx 1.42$ (middle), and \texttt{m12c} at $z \approx 2.28$ (bottom).
For each individual galaxy, the $z$-axis is defined to be aligned with the total angular momentum of all gas particles within $R_{90}$.
The `face-on' view is thus along the $z$-axis, and the `edge-on' view is along a direction perpendicular to the $z$-axis.
For each example, the left two columns of Fig.~\ref{fig:gala_def} display the face-on gas density map (upper left), edge-on gas density map (upper right), face-on stellar density map (bottom left), and face-on SFR map (bottom right).
The $R_{90}$ range is indicated by a white dashed circle in each map.
These images illustrate that \texttt{m12b} features a thin disk, \texttt{m12w} a thick disk, and \texttt{m12c} appears irregular.

\subsection{Calculation of metallicity gradients}\label{subsec:cal_metal_gra}

\begin{figure*}[ht!]

\begin{minipage}{0.33\linewidth}
 \centering
\subfigure{
 \label{fig:m12b_met}
 \centering
 \includegraphics[align=c,width=0.8\linewidth]{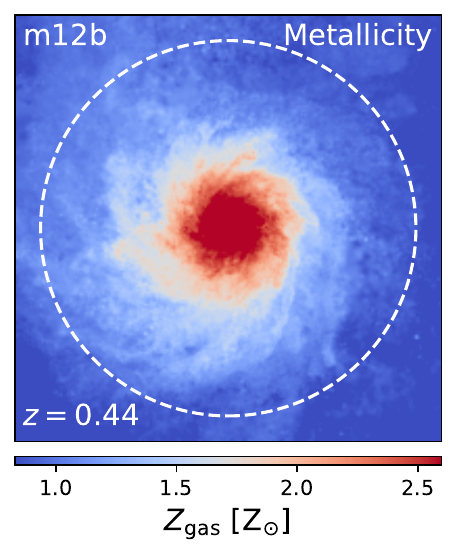}}
 \subfigure{
 \label{fig:m12b_gra}
 \centering
 \includegraphics[align=c,width=\linewidth]{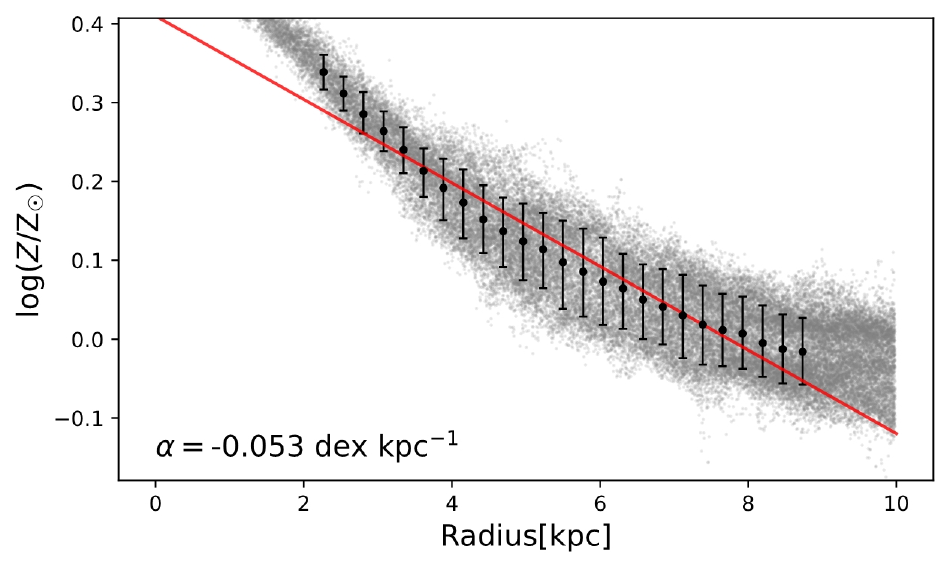}}
\end{minipage}
\begin{minipage}{0.33\linewidth}
 \centering
\subfigure{
 \label{fig:m12w_met}
 \centering
 \includegraphics[align=c,width=0.8\linewidth]{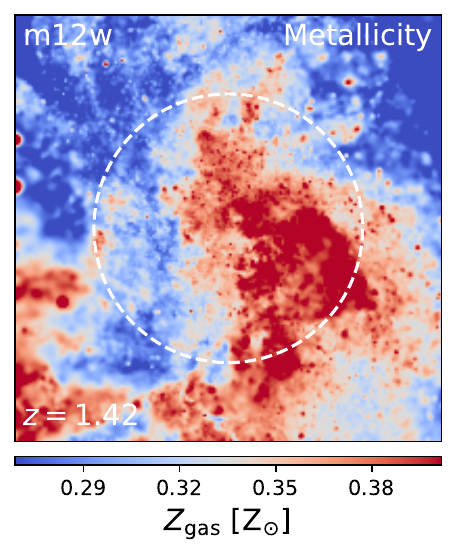}}
 \subfigure{
 \label{fig:m12w_gra}
 \centering
 \includegraphics[align=c,width=\linewidth]{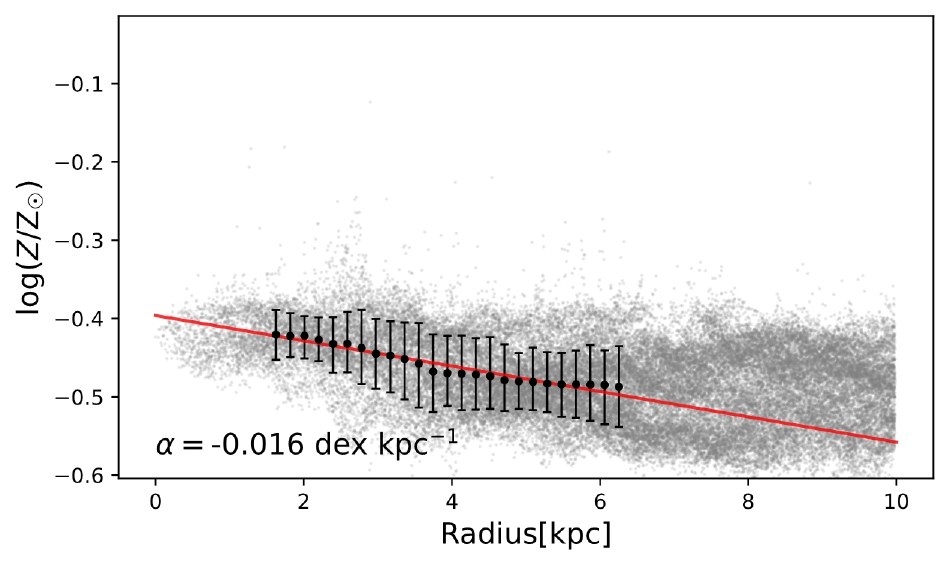}}
\end{minipage}
\begin{minipage}{0.33\linewidth}
 \centering
\subfigure{
 \label{fig:m12c_met}
 \centering
 \includegraphics[align=c,width=0.8\linewidth]{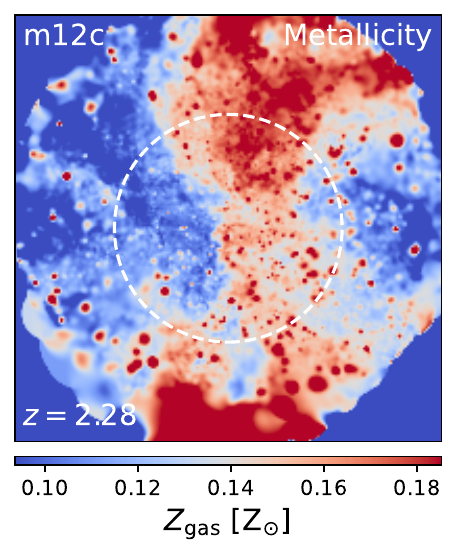}}
 \subfigure{
 \label{fig:m12c_gra}
 \centering
 \includegraphics[align=c,width=\linewidth]{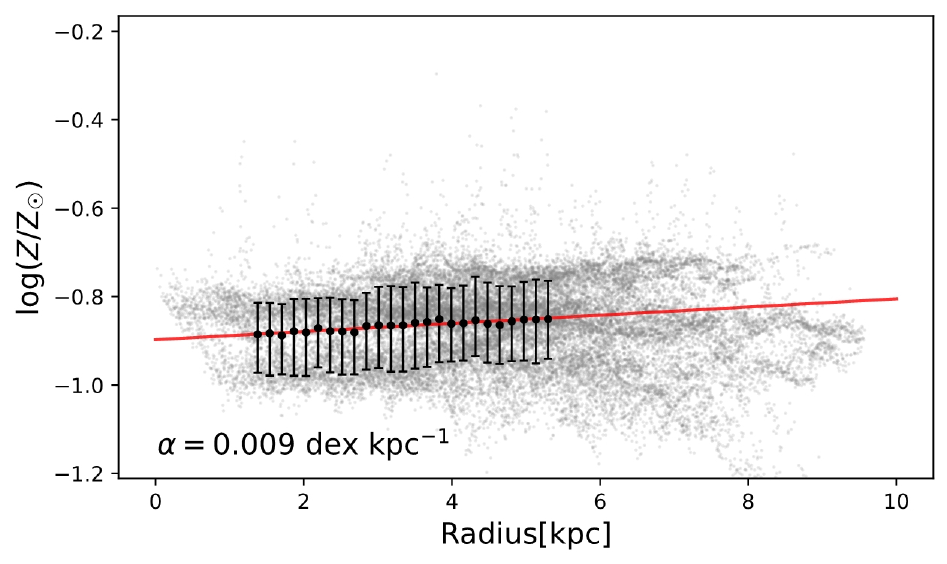}}
\end{minipage}

\caption{\emph{Top:} Face-on gas-phase metallicity maps for the example galaxies shown in Fig.~\ref{fig:gala_def}.
\emph{Bottom:} Radial gas metallicity profiles. The markers and error bars respectively represent the median and the 1-$\sigma$ uncertainty measurements of metallicity in each radial bin.
The red lines are the best linear fits according to Eq.~\ref{eq:alpha}, where $\alpha$ represents the slope of the metallicity gradient in dex$\cdot$kpc$^{-1}$.
Galaxy \texttt{m12b} at $z = 0.44$ exhibits a strong negative gradient; \texttt{m12w} at $z = 1.42$ shows a flat gradient; and \texttt{m12c} at $z = 2.28$ displays a positive gradient.
\label{fig:metal_gra}}
\end{figure*}

In the top panels of Fig.~\ref{fig:metal_gra}, we show the face-on metallicity distributions of the three galaxies in Fig.~\ref{fig:gala_def}.
The calculation is based on the total gas-phase metallicity provided for all 9 kinds of metals of all gas particles.
And the gas-phase metallicity is measured in each pixel (100 pc $\times$ 100 pc), and only pixels with gas mass density $\Sigma_{\rm g} \geq 10 {\rm M}_\odot\cdot {\rm pc}^{-2}$ are included in the metallicity gradient statistics.
This threshold represents the surface density of pixels above which there will be observationally detectable nebular emission lines, indicative of star formation \citep{Orr2018}.

Our analysis focuses on all particle data within the radius interval of 0.25-1$R_{90}$, following \cite{Ma2017}, which has shown that the metallicity gradients measured within a more inner range (0-2 kpc) yield qualitatively consistent results.
As shown by the red lines in the bottom panels of Fig.~\ref{fig:metal_gra}, the radial metallicity profile can be fitted with a linear function
\begin{equation}\label{eq:alpha}
    \log\frac{Z_{\rm g}}{Z_\odot}=\alpha R+\beta,
\end{equation}
where $\alpha$ is the metallicity gradient in $\mathrm{dex}\cdot\mathrm{kpc}^{-1}$.
Eq.~\ref{eq:alpha} provides a measure of the gradients of total metal in $d\log Z_{\rm g}/dR$.

In this study, positive gradient denotes $\alpha>0$, and we subdivide negative gradients ($\alpha<0$) into ``flat'' and ``strong negative'', where flat gradient denotes $-0.03<\alpha<0$, and strong negative gradient denotes $\alpha<-0.03$.

We note that for galaxies with a well-ordered rotation curve whose metallicity is concentrated at the center (e.g. \texttt{m12b} in Figs.~\ref{fig:gala_def} and \ref{fig:metal_gra}), the metallicity profile may be better modeled by a reciprocal ($d\log Z_{\rm g}/dR\sim 1/R$) rather than linear function, especially when considered over a larger radial distance (see Appendix~\ref{app:gala_changed_gra} for more details). Nonetheless, Eq.~\ref{eq:alpha} is still a good approximate for characterizing metallicity gradients.

\subsection{Kinematics}

We measure the kinematic properties for gas of these galaxies by mimicking the widely used long-slit spectroscopy technique following \cite{Ma2017}.
To begin with, we put a long slit with a width of 1 kpc at the center along the z-axis (edge-on) direction, as shown in the edge-on gas density images in Fig.~\ref{fig:gala_def}.
A one-dimensional velocity curve is then extracted along the direction of this slit, covering the range $-R_{90}< y <R_{90}$ and -0.5 kpc $< z <$ 0.5 kpc, with all gas particles therein considered in the calculation.
We measure the mass-weighted mean velocity of gas in each bin and the 1-$\sigma$ velocity dispersion of total velocity.
Three examples are shown in the right column of Fig.~\ref{fig:gala_def}, \texttt{m12b} at $z = 0.44$, \texttt{m12w} at $z = 1.42$, and \texttt{m12c} at $z = 2.28$, where black points represent the line-of-sight velocity and error bars indicate the velocity dispersion along the slit.

We fit the one-dimensional velocity curve with the following function
\begin{equation}\label{eq:vel}
    v(R)-v_0=v_{\rm c}\frac{2}{\pi}\arctan\frac{R}{R_{\rm t}},
\end{equation}
which originates from the simple disk model and similar to Eq.(1) in \cite{Ma2017} except that the peculiar velocity $v_0$ in the simulation box is placed on the left side.
Here $v_{\rm c}$ represents the asymptotic rotation velocity of the gas at large radius.
In the right panels of Fig.~\ref{fig:gala_def}, the red lines are the best-fits for the three examples.
On the top left corner of each panel, we provide $v_{\rm c}/\sigma$ for all the gas particles to indicate the disk formation of these galaxies, with $\sigma$ denoting the velocity dispersion of the line-of-sight velocity of all gas particles in the considered range.

We adopt $v_{\rm c}/\sigma\geq 1$ in identifying rotationally supported disk systems. Galaxies with $v_{\rm c}/\sigma< 1$ lack well-ordered rotation, and exhibit irregular morphology. And we define a threshold at $v_{\rm c}/\sigma = 3$, where galaxies with $v_{\rm c}/\sigma \geq 3$ are classified as thin disks, and galaxies with $1<v_{\rm c}/\sigma<3$ are classified as thick disks.

The velocity curves of \texttt{m12b} at $z=0.44$ can be well fitted by the arctan function, indicating that it possesses a well-ordered rotating thin disk, as evidenced by $v_{\rm c}/\sigma$ reaching 4.67.
Meanwhile, \texttt{m12w} at $z = 1.42$ can be roughly fitted, suggesting it possesses a thick disk with a $v_{\rm c}/\sigma$ of 2.49.
However, \texttt{m12c} at $z = 2.28$ returns a very poor fit, with a $v_{\rm c}/\sigma$ value of 0.05, indicating that it is an irregular galaxy.
These observations are consistent with the images in Fig.~\ref{fig:gala_def}.

In this study, we focus on galaxies with a $v_{\rm c}/\sigma$ value within the range of 0-$R_{90}$, as those with $v_{\rm c}/\sigma$ values beyond this range typically exhibit characteristics of irregular galaxies, which often show very unstable rotational velocities.

\section{Results} \label{sec:result}

\subsection{Metallicity gradients} \label{subsec:met_gra}

\begin{figure}[ht!]
 \centering
 \includegraphics[width=\linewidth]{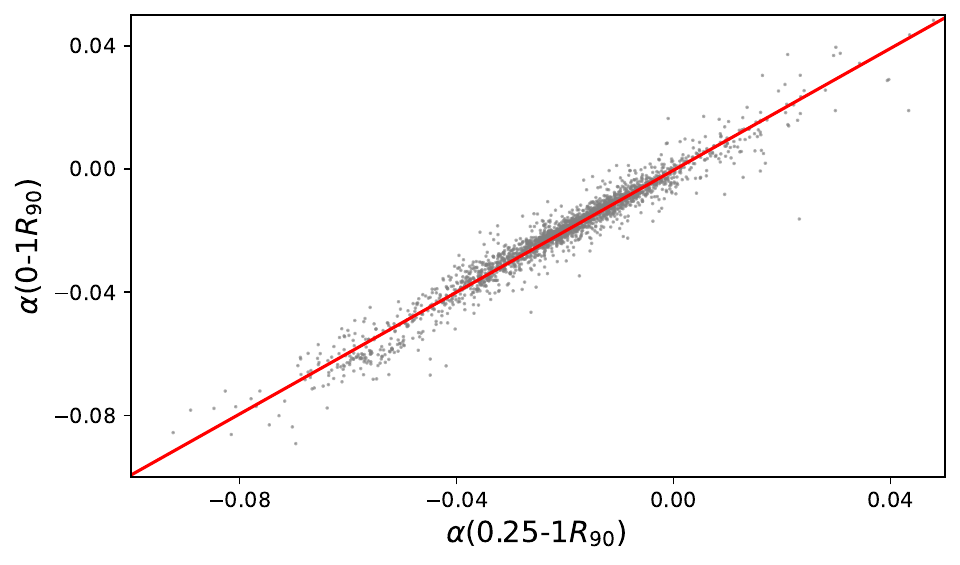}
 \caption{\emph{Comparison of gradients from 0-1$R_{90}$ and from 0.25-1$R_{90}$.} The two measurements are in excellent agreement for galaxies in our sample with a correlation coefficient of $k=0.989$ and no evidence for systematic differences.}
\label{fig:alpha_compare}
\end{figure}

Fig.~\ref{fig:metal_gra} displays the gas-phase metallicity maps and the radial metallicity distributions with linear fits for each galaxy snapshot in Fig.~\ref{fig:gala_def}. Galaxy \texttt{m12b} at $z = 0.44$ represents a galaxy in the later stages of evolution with a highly ordered rotational thin disk and a strong negative gradient with metals concentrated at the center.
Galaxy \texttt{m12w} at $z = 1.42$ indicates a mid-stage evolutionary galaxy with a thick disk, characterized by a flat gradient.
Galaxy \texttt{m12c} at $z = 2.28$ shows an irregular galaxy with a disordered velocity field, displaying a higher metal abundance in the outer regions, which is classified as a positive gradient.
See Appendix~\ref{app:gala_info} for the metallicity gradients of other galaxies in our study.
Previously, we mentioned that the results are similar across different regions.
In Fig.\ref{fig:alpha_compare}, we show that metallicity gradient measurements from 0-1$R_{90}$ and from 0.25-1$R_{90}$ are in excellent agreement with one another, with a correlation coefficient of $k=0.989$.

Intense outflows, rapid gas infall, and mergers in galaxies stir the ISM and drive galaxy-scale gas flows that mix metal-enriched or metal-poor gas.
These gas flows move at velocities of hundreds of km$\cdot$s$^{-1}$, which can reshape the spatial distribution of gas and metals in a very short time or cause changes in the galaxy's morphology \citep{Ma2017, Pandya2021}.
Therefore, simulations with "enhanced" feedback often exhibit flat metallicity gradients, while the absence of such feedback can lead to steeply negative gradients \citep{Gibson2013}. 
In the simulation \texttt{m12c} at $z = 2.28$, the perturbations are predominantly caused by a series of minor mergers, as observable in the images. These merger events across the galaxy mix metals in the gas, effectively reversing the gradient.
Meanwhile, for \texttt{m12w} at $z = 1.42$, the central starburst acts as the main source of feedback, driving the metal-enriched gas radially outward and consequently fostering the formation of a flat gradient.
On the other hand, in \texttt{m12b} at $z = 0.44$ (Fig.~\ref{fig:m12b_den}), the weak feedback (or lower SFR) produces less energy, which allows strong negative gradients to remain stable.
In galaxy evolution, different types of dynamical processes (such as mergers, starbursts and associated feedback, accretion, etc.), their intensities, and the interactions among them can lead to changes in their impact on the entire galaxy \citep{Mercado2021}.
This dynamic interaction influences not only galactic structure and star formation but also manifests in the observable variations in metallicity distribution across the galaxy.
Thanks to the high resolution of FIRE-2, we are able to resolve the movements of galactic winds on very small scales, the turbulence within the ISM, and its effects on radial mixing.

Fig.~\ref{fig:metal_gra} also illustrates how measurements of radial metallicity gradients in galaxies can be influenced by the choice of center.
In fact, when we choose starbursts as the center, flat and positive gradients are more easily observed.
It is easy to define the center for thin disk galaxies, while it is more challenging for thick disk or irregular galaxies.
As mentioned in \citet{Bellardini2021, Bellardini2022}, late-stage evolved thin disk galaxies exhibit minor variations in azimuthal angle;  however, early-stage galaxies experience additional azimuthal scatter due to gas flows.
Feedback mechanisms within the galaxy, such as localized star formation activities, influence azimuthal metal abundance variations and lead to a patchy distribution of metals (as shown in \texttt{m12c} at $z=2.28$ in Fig.\ref{fig:metal_gra}), resulting in observable an larger azimuthal scatter.
The impact of these dynamics is particularly significant in galaxies with active star formation or turbulent conditions, where localized events can rapidly alter the metal distribution.
These analyses, distinct from radial distributions, provide more information related to local conditions.
In this paper, we use the galaxy's center of mass as the center for measuring metallicity gradients, and we do not discuss other potential centers for measurement extensively.

In the following parts of this section
, we explore the relationships of metallicity gradient with stellar mass and sSFR in Section~\ref{subsec:ZtoM}.
In Section~\ref{subsec:Ztok}, we examine the correlation between metallicity gradient and the degree of rotational support.
Following that, in Section~\ref{subsec:Ztoz}, we discuss the dependence of metallicity gradient on redshift across all our samples, with a specific focus on \texttt{m12b}.
Finally, using \texttt{m12b} as an example, we investigate the formation of positive gradients in Section~\ref{subsec:positive}.

\subsection{Metallicity gradient versus stellar mass and sSFR}\label{subsec:ZtoM}

\begin{figure*}[ht!]
  \hspace{-0.2cm}
\subfigure{
 \label{fig:gala_alphatoM}
 \centering
 \includegraphics[align=c,width=0.47\linewidth]{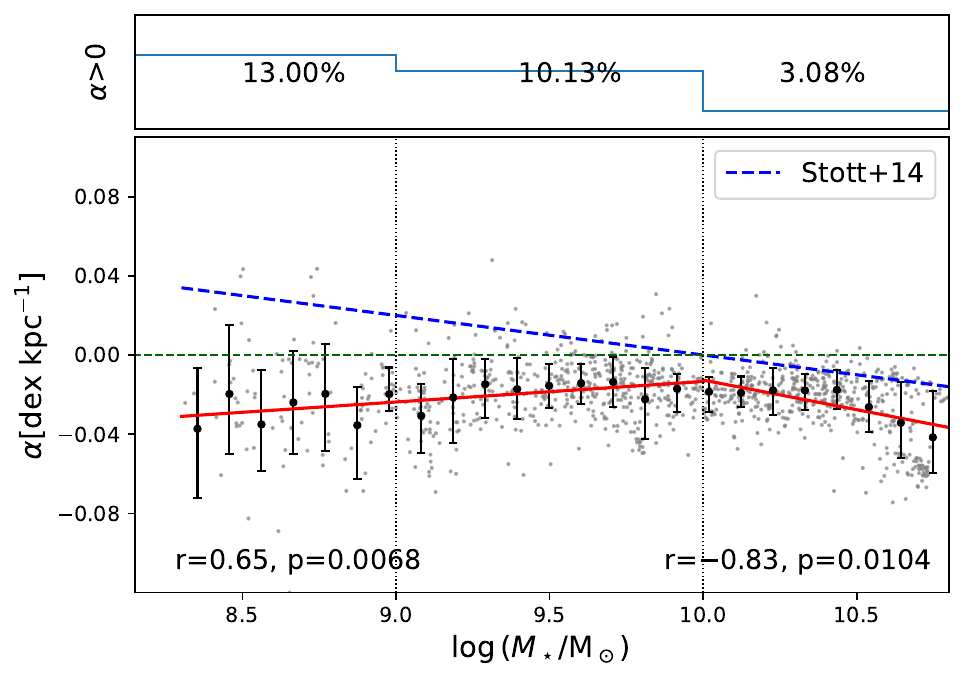}}
  \hspace{0.2cm}
\subfigure{
 \label{fig:gala_alphatoSFR}
 \centering
 \includegraphics[align=c,width=0.47\linewidth]{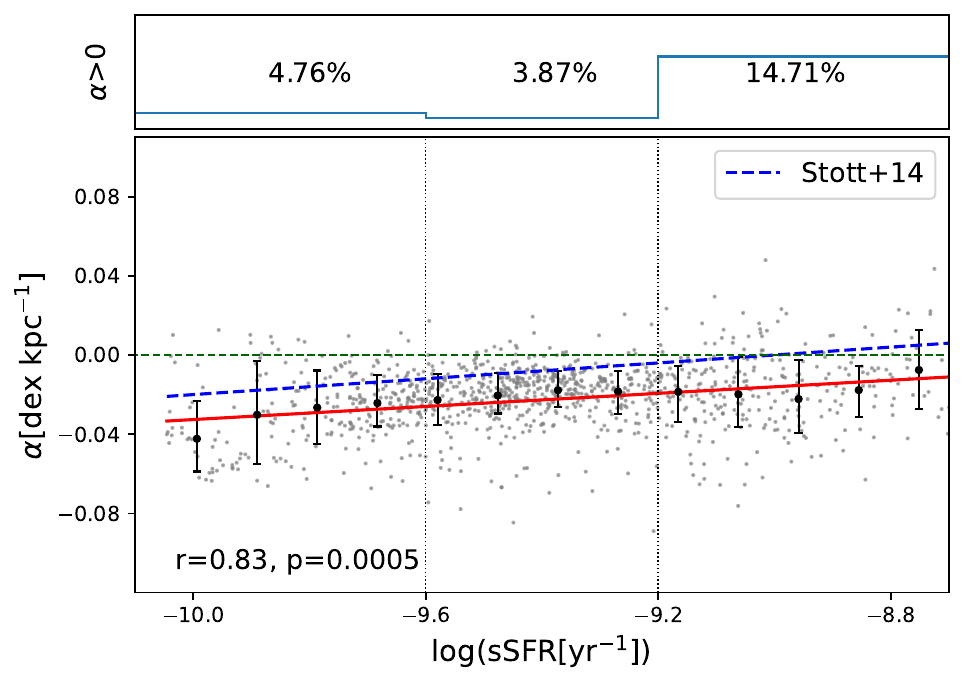}}
 \caption{\emph{Metallicity gradient versus stellar mass and sSFR.} sSFR are measured as the average formation rate of young stars over the past 200 Myrs. The red lines show the linear fit at $0.44<z<3$ from our FIRE-2 simulations. The blue dashed lines represent the linear fit to the observational data at $z = 0-2.5$ compiled by \citet{Stott2014}.
 Note that there is a positive correlation between stellar mass and redshift.
 Metallicity gradients show a weak negative correlation with stellar mass and a strong positive correlation with sSFR.
 Galaxies with low mass and high sSFR are more likely to have positive radial gradients.
 The text at the top indicates the proportion of positive gradients in different segments.
 }
\label{fig:gala_alphatoM-SFR}
\end{figure*}

We first examine the relationship between metallicity gradients, stellar mass, and sSFR. 
The SFR for a given epoch is measured as the average formation rate of young stars over the past 200 Myrs.
We display this relationship in Fig.~\ref{fig:gala_alphatoM-SFR} for all the galaxies across the redshift range of $z \sim 0.44-3$.
\citet{Ma2017} indicates that redshift changes do not significantly alter the underlying relationship between the metallicity gradient and these galaxy properties, so we do not distinguish between redshifts in this particular study.
Although we only use the \texttt{m12} series galaxies, where there is a direct correlation between the stellar mass and redshift, the relationship between gradient and stellar mass is not independent of redshift.
Nevertheless, due to the diverse evolutionary paths of different galaxies, the result still provides valuable insights into the relationships among them.
The red line represents the linear fits to the simulated data, while the blue dashed lines show the result from \cite{Stott2014} at redshifts $z \sim 0-2.5$.

Metallicity gradients do not exhibit a significant correlation with stellar mass, with a Pearson correlation coefficient of $r = -0.16$ and $p = 0.62$. 
However, as shown in the left panel of Fig.~\ref{fig:gala_alphatoM-SFR}, when the data are divided based on a threshold of $10^{10}M_\odot$, a notable correlation emerges: a positive correlation for $M_\ast < 10^{10}M_\odot$ ($r = 0.65$ and $p = 0.0068$) and a negative correlation for $M_\ast > 10^{10}M_\odot$ ($r = -0.83$ and $p = 0.0104$).
This trend mirrors the relationship between gradients and redshift, which, for the galaxies in our study, is inherently influenced by the connection between redshift and stellar mass.
Different to \cite{Stott2014}, our simulation provides a steeper fit, largely due to we use galaxies with overall higher mass (all $10^{12} M_\odot$).

Approximately $13\%$ of galaxies with stellar masses below $10^9 M_{\odot}$ exhibit positive gradients, and the proportion decrease to around $10\%$ for galaxies within the $10^9 - 10^{10} M_{\odot}$ range.
In other words, high-mass galaxies are less likely to exhibit positive gradients compared to lower-masses.
Low-mass galaxies have smaller galaxy scales, meaning they are more strongly influenced by the overall feedback, and making the mixing (turbulent) more effective.
This leads to the redistribution of metals on a galactic scale \citep{Muratov2015, Belfiore2017}, thereby more easily resulting in extreme gradients, like positive ones.
Additionally, toward larger mass at $M_\ast > 3 \times 10^{10} \rm{M}_\odot$, the metallicity gradient exhibits a sharp decline, attributed to that massive galaxies have already formed thin, stable disk structures \citep[see, e.g.,][for more on the emergence of thin disks and its connection to the end of bursty star formation in FIRE]{Stern2021_CGMFIRE, Hafen2022, Gurvich2023, Hopkins2023, McCluskey2024}.

We observe a strong positive correlation between gradients and sSFR, as reflected by a Pearson correlation coefficient of $r=0.83$ and $p=0.004$, showing an increased frequency of positive gradients at higher sSFR levels.
Higher sSFR drives stronger feedback, leading to more vigorous gas flows, which can more effectively promote the redistribution of metals.
Typically, strong star formation activity occurs in these galaxies that have more cold gas and are more unstable at high redshift, making them more susceptible to the resulting feedback.
That means, stellar winds driven by starbursts transport metal-enriched gas outward, or rapid accretion events (which drive galaxy activity) lead to intense metal redistribution \citep{Wang2019}, resulting in mixing and clumping, and forming positive gradients.
In stable, inactive galaxies, the accumulation of metals is more difficult to disrupt, resulting in strong negative gradients.

From our analysis of the FIRE-2 galaxy sample, we only see subtle dependence of metallicity gradient on stellar mass. 
Although, we caution that this analysis is performed on the same simulation galaxies over a wide redshift range across their mass assembly history, so there is an inevitable degeneracy between the effects that stellar mass and redshift plays in the evolution of gradients presented here and in Sect.~\ref{subsec:Ztoz}.
In contrast, the correlation between metallicity gradient and sSFR is stronger.
Additionally, for positive gradients, low-mass galaxies (which are more susceptible to feedback) and active galaxies (with stronger feedback) experience more vigorous gas motion in their interstellar medium, leading to metal redistribution on galaxy scale and resulting in the occurence of positive gradients.

\subsection{Metallicity gradient versus kinematic properties}\label{subsec:Ztok}

\begin{figure}[ht!]
 \centering
 \includegraphics[align=c,width=\linewidth]{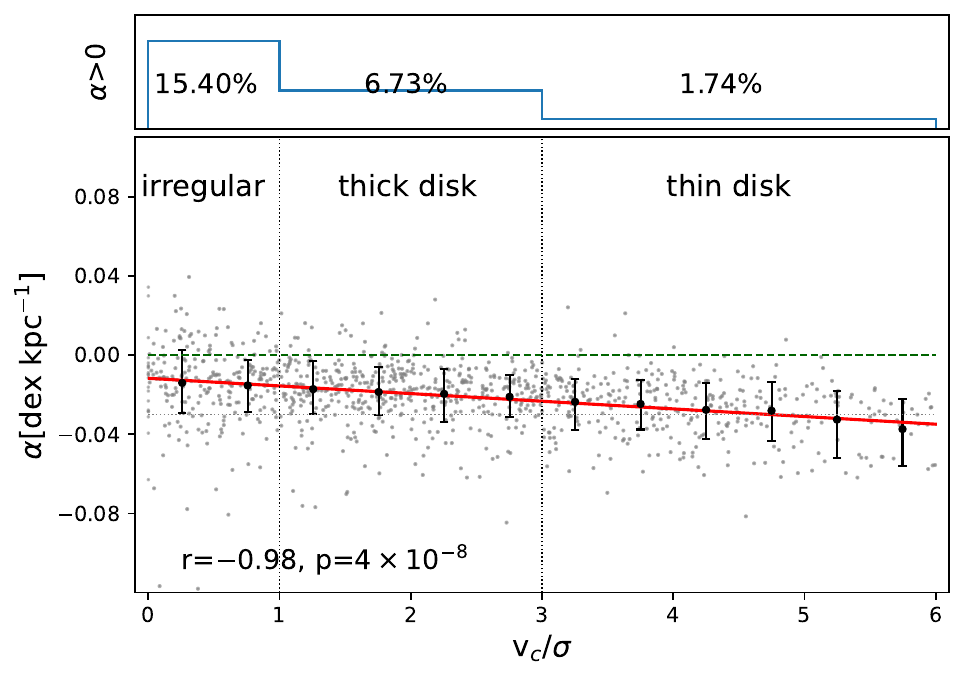}
 \caption{\emph{Metallicity gradient versus degree of rotational support for the ISM gas.}
 The red line represents the linear fit of our sample within the range of $v_{\rm c}/\sigma \sim 0-5$ at $0.44<z<3$ from our FIRE-2 simulations. 
 The metallicity gradient exhibits a extremely negative correlation with the degree of rotational support.
 Galaxies without rotationally supported disks are more likely to have flat and positive radial gradient. 
 The text at the top indicates the proportion of positive gradients in different segments.
 }
\label{fig:gala_alphatovs}
\end{figure}

In Fig.~\ref{fig:gala_alphatovs}, we illustrate the relationship between the gas-phase metallicity gradient and the degree of rotational support for gas, denoted by $v_{\rm c}/\sigma$.
Similar to the analysis for stellar mass and sSFR, the red line represents the linear fit to the simulated data.
Additionally, although the redshift of the \texttt{m12} series galaxies correlates with disk formation, the insights derived from the differences among various galaxies provide substantial reference value for the relationship.

Within the range of $v_{\rm c}/\sigma\sim$ 0-6,  the gradients exhibit a very strong negative correlation with the degree of rotational support, with a Pearson correlation coefficient of $r=-0.97$ and $p=3\times10^{-7}$, suggesting that galaxies with thinner disks tend to have stronger negative metallicity gradients.
For convenience, all galaxies are classified into three regions according to their gradient and $v_{\rm c}/\sigma$ as shown by the gray dotted lines in Fig.~\ref{fig:gala_alphatovs}.

For $0<v_{\rm c}/\sigma<1$, the internal dynamics of galaxies are unstable, making them more susceptible to disturbances caused by strong feedback-driven gas flows. 
This leads to large-scale metal redistribution within the galaxy, resulting in the positive gradients we observe.
As a result, the proportion of positive gradients is relatively high ($\sim 15\%$) in these irregular galaxies.
Galaxies with rotationally supported disks have a higher likelihood of exhibiting strong negative gradients.
Meanwhile, the proportion of positive gradients is decreasing as well, reducing to $\sim 7\%$ in thick disks, and only $\sim 2\%$ in thin disks.
In Fig.~\ref{fig:svtoz}, the redshift evolution of sSFR and $v_c/\sigma$ shows that as galaxies transition from a disordered state to a rotationally supported disk, their sSFR decreases, reflecting a reduction in strong feedback driven by earlier starburst activity. 
The strong correlation between the metallicity gradient and $v_c/\sigma$ further indicates that the formation of a rotationally supported disk significantly suppresses metal mixing processes driven by gas dynamics, including gas inflows, outflows and turbulent mixing \citep{Graf2024arXiv}.
Similar to FIRE-1 galaxies in \cite{Ma2017}, almost all galaxies with strong negative gradients are likely to be rotationally supported, but not vice versa.
Stable disk galaxies inhibit these powerful gas flows across the entire galaxy, while localized, orderly gas exchanges gradually maintain stability and promote the formation of strong negative gradients.

\subsection{Metallicity gradient versus redshift}\label{subsec:Ztoz}

\begin{figure*}[ht!]
 \centering
 \includegraphics[width=0.91\linewidth]{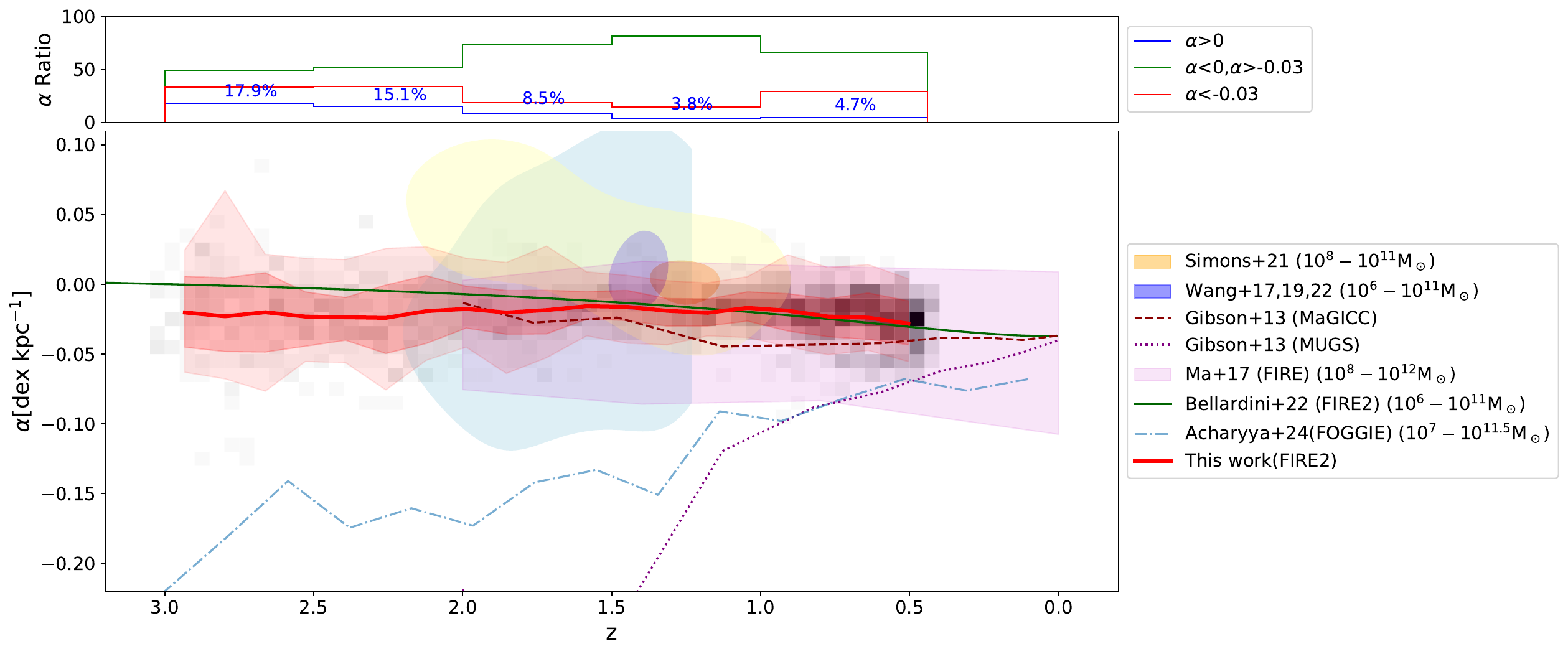}
 \caption{\emph{Metallicity gradient versus redshift.} 
 The grayscale colormap represent the metallicity gradients of our FIRE-2 simulated galaxies measured within 0.25-1$R_{90}$.
 The red line and shaded region indicate the median, 1-$\sigma$ and 2-$\sigma$ spread of our measurements.
 In comparison, we display the results from some recent high-$z$ spatially resolved spectroscopy from HST and JWST with the incidence rate of positive gradients \citep[][$16\%$, 3-$\sigma$]{Simons2021}, \citep[][$10-20\%$]{Wang2017, Wang2019, Wang2020, Wang2022}, and data from simulations, i.e., MUGS and MaGICC \citep{Gibson2013}, FIRE-1 \citep[][$13\%$]{Ma2017}, FIRE-2 \citep{Bellardini2022}, and FOGGIE \citep{Acharyya2024arXiv}.
 Our results show a flatter trend, with a slight increase at $z>1.5$, following a sharp decline at $z<1$. 
 An analysis of the proportion of different gradient magnitudes shows that as redshift decreases, the proportion of positive and strong negative gradients decrease, while the proportion of flat gradients increases, reaching its highest at $1<z<1.5$, where strong negative gradients are minimal.
 At $z<1$, the occurrence of strong negative gradients dramatically increases.
 The panel at the top shows the proportions of positive gradients in different redshift ranges.}
\label{fig:gala_alphatoz}
\end{figure*}

In Fig.~\ref{fig:gala_alphatoz}, the metallicity gradients of our simulated galaxies at $z \sim 0.44-3$ are represented as gray region, while the red line indicates the average for each redshift bin.
The majority of the gradients for our galaxies fall within the range from -0.08 to 0.04.
The FIRE-2 galaxies exhibit consistency with the results from the FIRE galaxies \citep{Ma2017} and are similar to those presented by MaGICC (enhanced feedback in \citep{Gibson2013}).
Compared to observational results \citep{Simons2021, Wang2017, Wang2019, Wang2020, Wang2022}, FIRE-2's is noticeably lower at same redshifts, partly due to the generally larger mass of our galaxies, which leads to similar evolution.
Nevertheless, FIRE-2 still shows roughly similar results to observations within the 1-sigma range.
All results ultimately converge between -0.05 and -0.1, suggesting that galaxies evolving into thin disks should have metallicity gradients within this range.

Throughout the chemo-structural evolution of these galaxies, positive metallicity gradients account for a total of $7\%$, a proportion similar to that reported in \cite{Montero2016, Carton2018, Wang2020}.
This proportion also aligns with the trend reflecting a transition from disorder to order \citep{Zhuang2019}.
In contrast, \citet{Simons2021} reported a significantly higher proportion of $29\%$ for 
their emission-line selected galaxy sample at $0.6<z<2.6$.
Yet we notice that only X-ray bright active galactic nuclei (AGNs) are removed from their galaxy sample, and AGN ionization can strongly increase the \ionf{O}{iii}$\lambda$5007/\ionf{O}{ii}$\lambda\lambda$3727,3730 line flux ratio in galaxy centers, 
resulting in artificial positive metallicity gradients.
From the numerical simulation side,
\citet{Gibson2013} conducted two different sets of simulations; the conservative feedback model (MUGS) showed that galaxies establish very strong negative metallicity gradients at high redshifts, which gradually flatten as the galaxies grow.
A similar phenomenon occurs in \citet{Acharyya2024arXiv}, where they state that the feedback mechanisms employed in the current FOGGIE simulations are overall underpowered.
While the "enhanced" feedback models maintained flat metallicity gradients at high redshifts and evolved to become steeper over time.
Based on the feedback model and the burstiness of star formation in the FIRE-2 simulations, our results show a wide scatter in metallicity gradients, indicating that these average metallicity gradients steepen with the evolution of redshift \citep[See also:][]{Ma2017}.
Although the trends toward steeper gradients at lower redshifts are similar, the average gradients we observe at higher redshifts not as flat as those reported by \cite{Bellardini2022}.
The gradient calculations for these high-redshift galaxies are influenced by various factors, including the computational ranges used in this paper ($0-0.25 R_{90}$).
Furthermore, the inclusion of ELVIS LG-like galaxies in their calculations may also contribute to the observed discrepancies.

The proportion of positive metallicity gradients decreases with lower redshifts, as shown in Fig.~\ref{fig:gala_alphatoz}.
This trend is also reflected in the relationships between metallicity gradients and mass, sSFR, and $v_c/\sigma$ discussed in previous sections, where positive metallicity gradients change accordingly.
We are not claiming that these are the sole causes, but they likely work together to promote the variation of positive gradients with redshift.
Additionally, the evolution of sSFR and $v_c/\sigma$ with redshift, shown in Fig.~\ref{fig:svtoz}, further supports the idea that various evolving variables with redshift can influence the formation of positive gradients.
Moreover, in the plots of metallicity gradients and $v_c/\sigma$ evolution with redshift (which also applies to mass), a distinct steep change appears on the right side, suggesting that once galaxies evolve into thin disks, the accumulation of metals across the galaxy may be further enhanced.


\begin{figure*}[ht!]
 \centering
 \includegraphics[width=\linewidth]{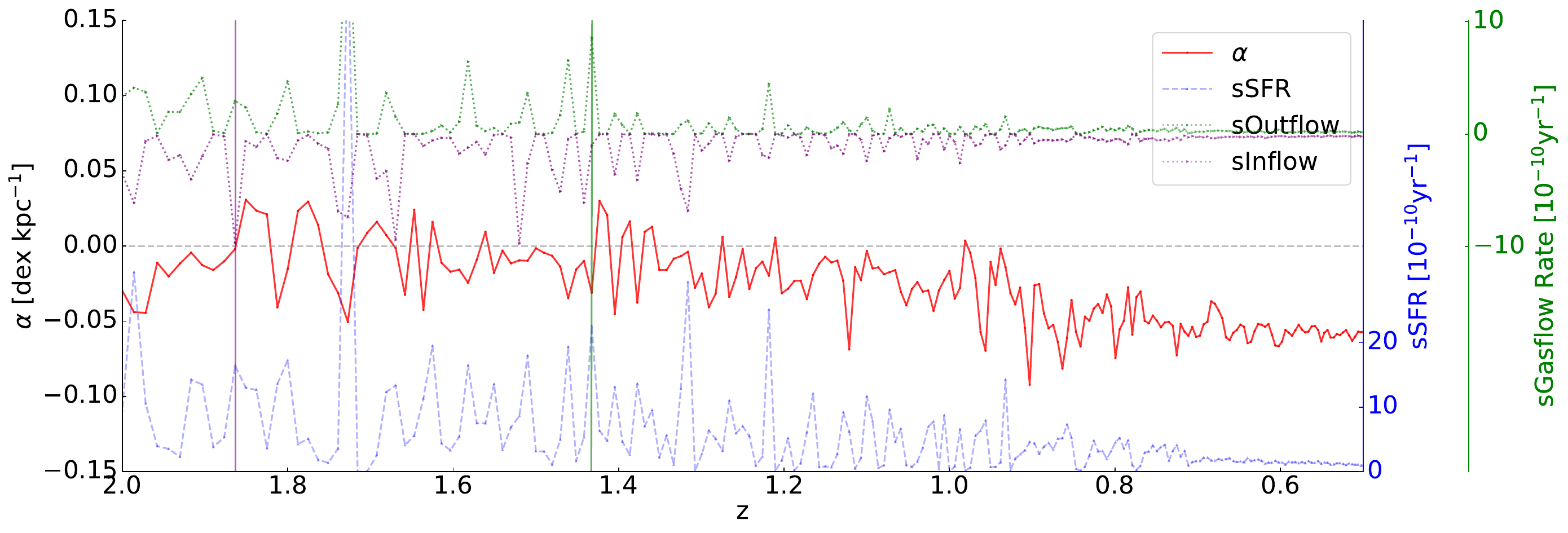}
 \caption{\emph{Evolution of metallicity gradient, sSFR and specific-gasflow rate in galaxy \texttt{\rm{m12b}}.}
 Here, the red line represents the metallicity gradient, and the blue line represents the sSFR, green dotted line shows the specific-outflow rate, purple dotted line shows the specific-outflow rate across $z \sim 0.5-2$.
 The vertical green and purple lines respectively mark the outflow and inflow of gas preceding the positive metallicity gradient.
 Generally, we refer to these two phenomena as metal-enriched gas outflow and metal-poor gas inflow.
}
\label{fig:m12b_alphatoz}
\end{figure*}

In Fig.~\ref{fig:m12b_alphatoz}, we display galaxy \texttt{m12b}, including gradients in red, sSFR over the last 10 Myrs in blue, and the specific-outflow rate in green, specific-inflow rate in purple (will be introduced in the Section~\ref{subsec:positive}) across $z \sim 0.5-2$.
The proportion of positive metallicity gradients of \texttt{m12b} is $\sim9.3\%$, which is roughly consistent with observed across all galaxies.
The evolution of the galaxies is consistent with the overall observations.
Moreover, their evolution with redshift also reflects this consistency throughout the entire galaxy evolution process, where the occurrence of positive gradients is typically associated with the presence of strong feedback.
We also present the evolution of all galaxies in Appendix~\ref{app:gala_evol}. 
Although they show consistency in their evolutionary trends, there is considerable variation between the galaxies, indicating that the metals in galaxies are highly sensitive to various internal feedback mechanisms, which leads to the observed differences in gradients.

We observe a clear evolution for these Milky Way-mass progenitors galaxies: 
at redshifts $z>1.5$, galaxies typically exhibit bursty star formation activity, leading to extremely strong outflows \citep{Muratov2015, Muratov2017, Angles2017, Pandya2021}, which facilitates radial metal mixing in ISM, manifesting in flat and positive gradients.
Around $z\sim1.5$, the average gradient peaks as the internal dynamics and structure of galaxies stabilize.
The relative stability in the physical structure of galaxies supports the formation and maintenance of more stable gradients.
From $z\sim1.5$ to $1$, as the intensity of stellar activities declines, the impact of feedback mechanisms weakens, and rotating disks prevent strong feedback that spans the entire galactic scale, leading to weaker radial metal mixing or turbulence in the ISM \citep{Graf2024, Graf2024arXiv}.
Consequently, there is a noticeable steepening in the average metallicity gradients across galaxies.
Then $z\sim1-0.7$, galaxies transition further to rotational disks, and the presence of stronger ordered rotating thin disks inhibits radial mixing in the ISM.
The SFR decreases from bursty to time-steady \citep{Gurvich2023}, and galactic-scale gas outflows or turbulence are suppressed; the disordered accretion from CGM shifts to an ordered form (namely, virialization of the inner CGM) \citep{Stern2021_CGMFIRE}, leading to further steepening in the average metallicity gradients.
Below $z<0.7$, dynamics is dominated by weak ``fountain flows'' \citep{Angles2017} and ordered accretion from the CGM.
The effective radial redistribution of metals is nearly prevented, and as metals continue to enrich in thin disk galaxies, stabilizing metallicity at steep negative values around -0.08.

\subsection{Positive metallicity gradient and gas flow}\label{subsec:positive}

After discussing the overall evolution of galaxies, we aim to briefly explore the simple relationship between gas flows and metallicity gradients using Fig.~\ref{fig:m12b_alphatoz}. 
Here we follow \citet{Faucher2011}, \citet{Muratov2015} and \citet{Ma2017} to calculate the gas outflow rate as
\begin{equation}\label{eq:flow}
    \frac{\partial M}{\partial t} = \frac{1}{L}\sum m\frac{\boldsymbol{v}\cdot\boldsymbol{r}}{\lvert \boldsymbol{r}\rvert}
\end{equation}
and the specific-outflow rate as $\dot{M}/M_\ast$.
In this work, we simply consider all gas particles with a radial velocity $v_r = \boldsymbol{v} \cdot \boldsymbol{r}/\lvert \boldsymbol{r}\rvert > 0$ as outflow gas with in central $L = 0.25R_{90}$ for the inner region.
Additionally, we attempted to explore its direct relationship with the metallicity gradient, but found no significant correlation.

As indicated by the vertical green solid line in the figure, it is clear that immediately after the gas outflow, there is a notable shift in the metallicity gradient, which transitions into a positive gradient. This change coincides with the peak of sSFR, suggesting that the star formation activity plays a crucial role in influencing the gas flows in the galaxy, thereby redistributing metals. 
Also, as shown by the vertical purple solid line, the positive metallicity gradient follows the gas inflow. This indicates that the inflow of gas also plays a role in shaping the metallicity distribution. 
These gases with different metallicities cause large-scale metal redistribution during their flow, directly leading to the diversity of metallicity gradients. 
Typically, the first phenomenon is called metal-enriched gas outflow, where metals are expelled to outskirts of galaxy; while the second is known as metal-poor gas inflow, where gas with low metal content flows into the center of galaxy \citep{Wang2022}.

Additionally, it is worth noting that the intensity of gas flow is closely related to the specific star formation rate (sSFR), with both showing similar trends in their evolution with redshift. Typically, strong gas flows—whether inflows or outflows—are observed alongside peaks in the sSFR, indicating periods of heightened galaxy activity. These moments of intense gas movement are crucial for shaping the galaxy's metal distribution and star formation activity, as they reflect the underlying feedback processes and energy exchanges within the galaxy.

In our study of positive gradients, many gradients are found quite flat, the reasons of which are similar to those for positive gradients.
In Fig.~\ref{fig:m12b_alphatoz} we could see that the weakening of strong feedback and the suppression of galaxy-scale gas flows make many gas flows too weak for positive gradients, yet sufficient for flat gradients at $z\sim1.0$.
We are not suggesting that all flat gradients originate in the same way as positive gradients.
Some may arise from strong turbulence within the galaxy, which merely mixes metals (rather than altering the large-scale metal distribution), leading to the formation of flat gradients.

\section{Conclusions} \label{sec:conclusion}

In this work, we use 8 high-resolution cosmological zoom-in simulations from the core suite (all of them are \texttt{m12} galaxies) of the FIRE-2 project to examine the correlation between gas-phase metallicity and other galaxy properties, thereby inferring the influences acting upon them.
These samples cover a redshift range of $z \sim 0.44-3$ and stellar masses of $M_{\ast} \sim 10^8-10^{11}$ M$_{\odot}$, with 281 snapshots available for each galaxy.
Thanks to the high resolution and detailed physical processes modeled by FIRE-2, we can observe the evolution of galaxies and the effect of feedback mechanisms on the spatial distribution of gas-phase metallicity.

We investigated the relationships among galaxy metallicity gradient, stellar mass, sSFR, and the degree of rotational support across the entire redshift range of $0.4<z<3$.
Across all galaxies, positive gradients make up about $7\%$ of the total.
\begin{itemize}
    \item There is no correlation between stellar mass and gradient, but when the low-to-mid mass and high-mass ranges are separated, different correlation can be found.The average gradient flattens with mass and then decreases rapidly.Nevertheless, the proportion of positive metallicity gradients observed remains higher in low-mass galaxies and decreases with increasing mass. In other words, low-mass galaxies have smaller scales, which makes them more affected by feedback, leading to the formation of positive gradients.
    \item There is a strong positive correlation between metallicity gradients and sSFR, with more active galaxies showing flatter average metallicity gradients. That is to say, stronger feedback results in greater impacts on galaxies, where gas flows driven by starbursts can rapidly reshape the spatial distribution of metal in galaxy, leading to the formation of positive gradients.
    \item The metallicity gradient of galaxies exhibits a strong negative correlation with their rotational support. Based on the $v/\sigma$ values, they could be classified into the following three types: irregular galaxies, thick disk galaxies with preliminary rotational properties, and thin disk galaxies with rotation-dominated disk. Rotation-dominated galaxies show predominantly strong negative gradients. Thick disk galaxies tend to have flat gradients, with only a few showing positive ones. Irregular galaxies exhibit the highest proportion of positive gradients. In galaxies without a disk, metals are more easily scattered by feedback. Strong feedback-driven gas flows can effectively transport metal-enriched gas to the outskirts, resulting in positive gradients. In contrast, galaxies with stable disks can suppress such powerful galactic outflows, causing metals to circulate and accumulate in fountain-like gas flows, leading to the formation of strong negative gradients.
\end{itemize}

Since all our galaxies are selected to have a halo mass of $\sim 10^{12}\rm M_\odot$ at $z\approx 0$, they exhibit similar characteristics.
Over long timescales, the metallicity gradients of entire samples show a downward trend from cosmic noon to lower redshifts.
At high redshifts ($z \gtrsim 1.5$), galaxies exhibit intense star formation activity. Strong enough feedback effects drive gas flows on the galaxy scale, facilitating the expansion of metals to the outer regions of the galaxy, resulting in a dominant positive gradient.
At medium redshifts ($z \sim 1-1.5$), galactic activities gradually decrease, and galaxies transition from disordered states to quasi-stable thick rotating disks.
Concurrently, the influence of gas flows diminishes to sub-galactic scales, and the virialization of the CGM occurs \citep{Stern2021_CGMFIRE}.
Metals begin to accumulate, leading to decreased metallicity gradients, with the proportion of positive gradients dropping.
At low redshift $z < 1$, galaxies are in a stable thin disk state, with reduced radial mixing in the ISM, gas cycling in fountain-like flows supports the maintenance of stable, strong negative gradients \citep{Graf2024arXiv}.
The distinct characteristics of different galaxies determine their unique evolutionary traits. Nevertheless, they exhibit consistency in the overall direction of their evolution.

This work focuses on the overall properties of galaxies and explores the reasons for their association with metallicity gradients.
We did not delve into the specific directions of gas flows, which are related to more detailed feedback mechanisms within galaxies.
Further studies on gas and metal flow across galaxies may be presented in future work.

\begin{acknowledgments}
We thank the anonymous referee for a very constructive report that helps improve the clarity of this paper.
This work is supported by the National Natural Science Foundation of China (grant 12373009), the CAS Project for Young Scientists in Basic Research Grant No. YSBR-062, the Fundamental Research Funds for the Central Universities, and the China Manned Space Program with grant no. CMS-CSST-2025-A06.
XW acknowledges the support by the Xiaomi Young Talents Program, and the work carried out, in part, at the Swinburne University of Technology, sponsored by the ACAMAR visiting fellowship.
AW received support from NSF, via CAREER award AST-2045928 and grant AST-2107772, and HST grant GO-16273 from STScI.
CAFG was supported by NSF through grants AST-2108230 and AST-2307327; by NASA through grant 21-ATP21-0036; and by STScI through grant JWST-AR-03252.001-A.
\end{acknowledgments}

\vspace{5mm}

\appendix

\section{Gradient Radial Variations}\label{app:gala_changed_gra}

\begin{figure*}[ht!]
\subfigure{
 \label{fig:galab_alphator}
 \centering
 \includegraphics[align=c,width=0.48\linewidth]{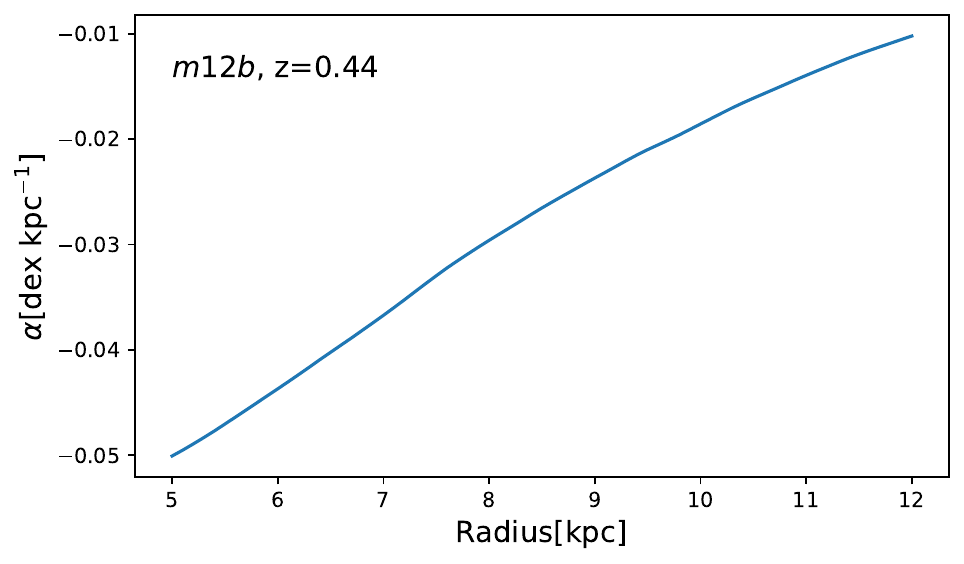}}
\subfigure{
 \label{fig:galab_mentor}
 \centering
 \includegraphics[align=c,width=0.5\linewidth]{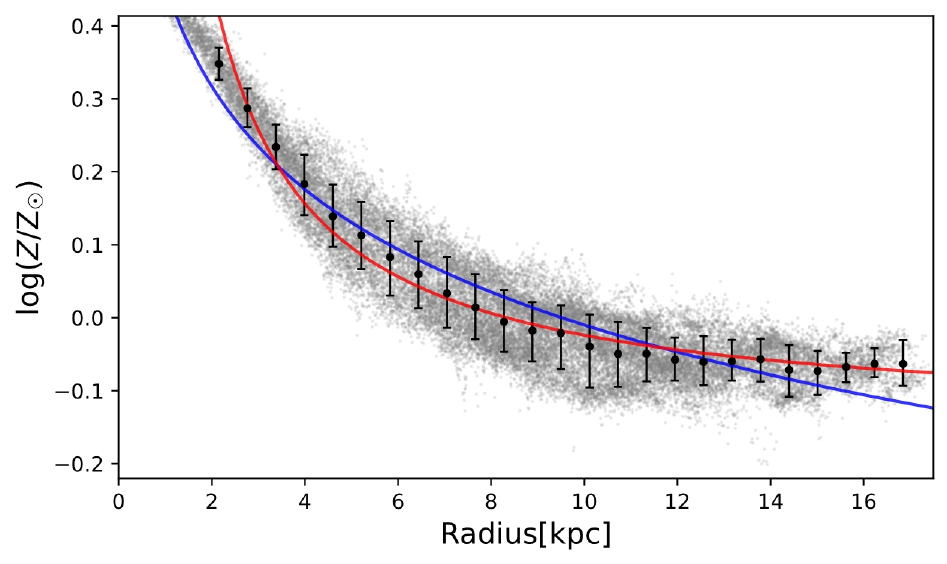}}
\subfigure{
 \label{fig:galac_alphator}
 \centering
 \includegraphics[align=c,width=0.48\linewidth]{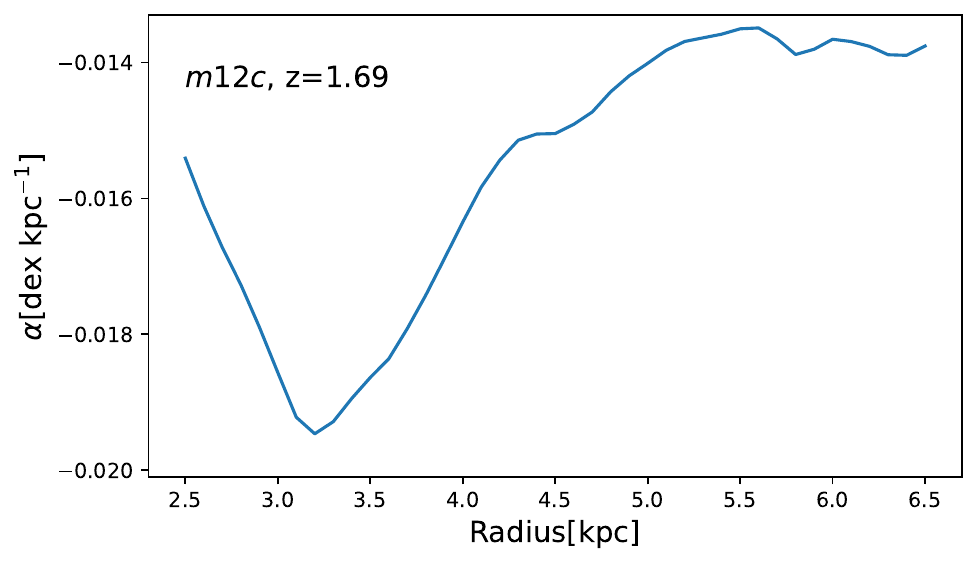}}
\subfigure{
 \label{fig:galac_mentor}
 \centering
 \includegraphics[align=c,width=0.5\linewidth]{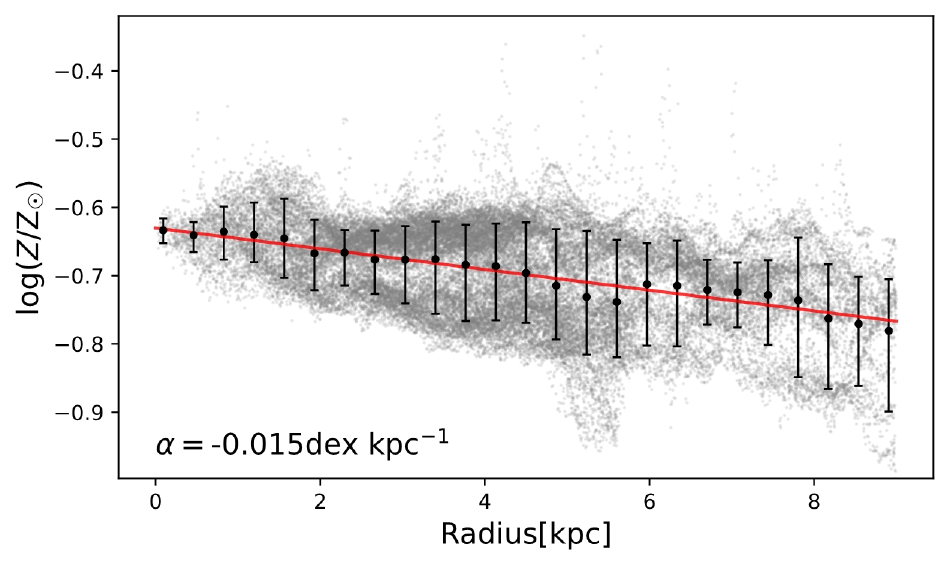}}
\caption{\emph{Left:} Galaxies \texttt{m12b} at redshift $z = 0.44$ and \texttt{m12c} at redshift $z = 1.69$, displaying their gas-phase metallicity gradients versus radius.
\emph{Right:} the metallicity of \texttt{m12b} fitted using a reciprocal (red) and a logarithmic (blue) function, and \texttt{m12c} using a linear (red) function.
The metallicity gradient becomes steeper towards the inner part of the galaxy.
We calculate the gradient over a range of $\pm$5 kpc, totaling a 10 kpc span, around 5 - 12 kpc from the center, observing a maximum difference of about 17-fold between the highest and lowest values.
For \texttt{m12c} it's a 5 kpc span ($\pm$2.5 kpc), around 2.5 - 9 kpc from the center, with the maximum difference being only 1.5-fold.}
\label{fig:gala_alphator-gala_met}
\end{figure*}

On the upper left of Fig.~\ref{fig:gala_alphator-gala_met}, we illustrate the relationship between gas-phase metallicity gradient and radius of galaxy \texttt{m12b} at redshift $z = 0.44$.
Here we consider a range from 5 to 12 kpc, and every region is composed of $\pm$5 kpc, totaling a 10 kpc span for the calculation of the gradient.
The maximum difference of the gradient, defined as the absolute change from its highest to its lowest value, is about 17-fold.
However, in \texttt{m12c} at $z = 1.69$, where each region spans $\pm$2.5 kpc from 2.5 to 6.5 kpc, different results are observed.
Within whis range, the maximum difference found is only 1.5-fold.

In a disk galaxy, the observed gradient tends to become more steeper while closer to the interior, which is also mentioned in \cite{Bresolin2012, Bresolin2017}.
For various metal elements, their gas-phase metallicity exhibit such a distribution pattern, and similarly, the stellar metallicity also display a comparable distribution \citep{Bellardini2021, Bellardini2022}.
From this we could infer that in the disk galaxies, there is not only higher metallicity, but also the change of metallicity could not simply be described using a linear function as is shown on the right of Fig.~\ref{fig:gala_alphator-gala_met}.
We used reciprocal and logarithmic function for the fitting, as shown by the red and blue line:
\begin{equation}\label{eq:red}
    Red: \log\frac{Z_g}{Z_\odot}=-0.14+1.20\frac{1}{R\ (\rm{kpc})},
\end{equation}
\begin{equation}\label{eq:blue}
    Blue: \log\frac{Z_g}{Z_\odot}=0.46-0.20\ln R\ (\rm{kpc}).
\end{equation}
Here, the reciprocal (red) function provides a better fit than the logarithmic (blue) function.
This phenomenon may be related to the dynamics, as in our sample, it is often found to occur in disk galaxies.
Additionally, it may reflect the way in which the galaxy formed.

However, for high redshift galaxies or those that have not formed a thin disk, we could not observe phenomena like this.
The change in the gradient of \texttt{m12c} is very small, as shown in Fig.~\ref{fig:galac_alphator}.
This means that for the high-redshift galaxies, especially for those with flat and positive gradients, a linear fitting usually meets our requirements. 

\begin{figure*}[ht!]
  \hspace{-0.2cm}
\subfigure{
 \label{fig:stoz}
 \centering
 \includegraphics[align=c,width=0.47\linewidth]{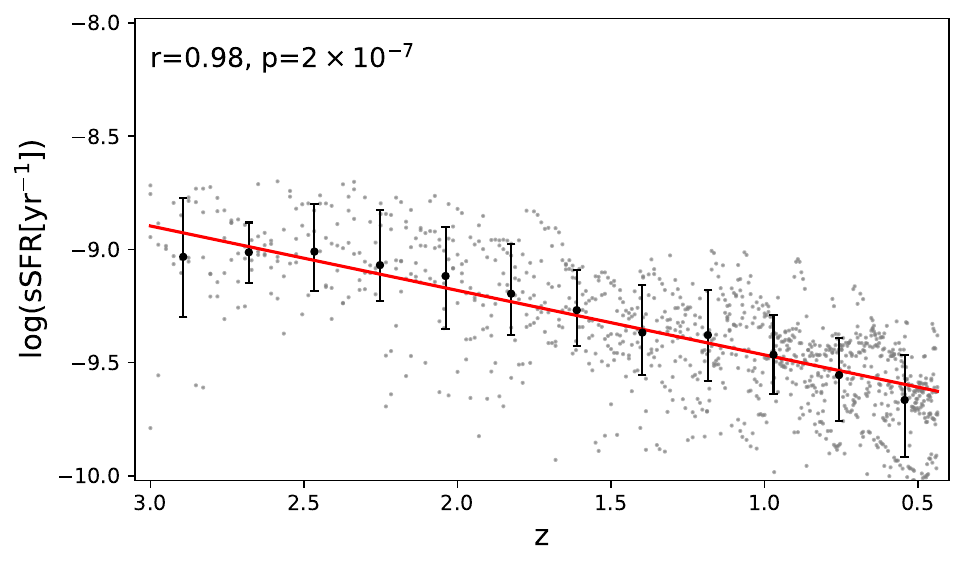}}
  \hspace{0.2cm}
\subfigure{
 \label{fig:vtoz}
 \centering
 \includegraphics[align=c,width=0.47\linewidth]{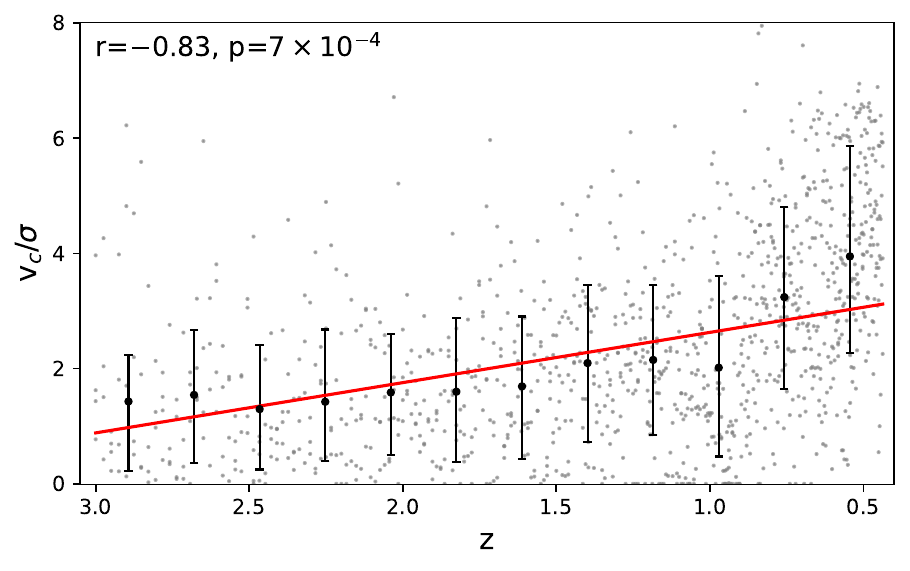}}
 \caption{\emph{sSFR and $v_c/\sigma$ versus redshift. Both show correlation with redshift. Although the linear fit is slightly inadequate, the evolutionary trend is still evident in this work.}
 }
\label{fig:svtoz}
\end{figure*}

\section{Galaxy Information}\label{app:gala_info}

In this section, we present in Table~\ref{tab:galadata} the properties of selected galaxy samples, including stellar mass, sSFR over $200\,\rm Myrs$, $R_{90}$, gas-phase metallicity gradients measured from $0.25-1R_{90}$ (denoted as $\alpha$), and kinematic properties ($v_{\rm c}/\sigma$).

\section{Galaxy Evolution}\label{app:gala_evol}

In this section, we present the evolution of sSFR and $v_c/\sigma$ to $z$ in Fig.\ref{fig:svtoz}. The sSFR gradually decreases with redshift, indicating that galaxies are becoming more stable, and the relative strong feedback across the entire galaxy weakens. On the other hand, the morphology of galaxies becomes more stable, forming a disk-like structure.

\begin{figure*}[ht!]
 \centering
 \includegraphics[width=0.8\linewidth]{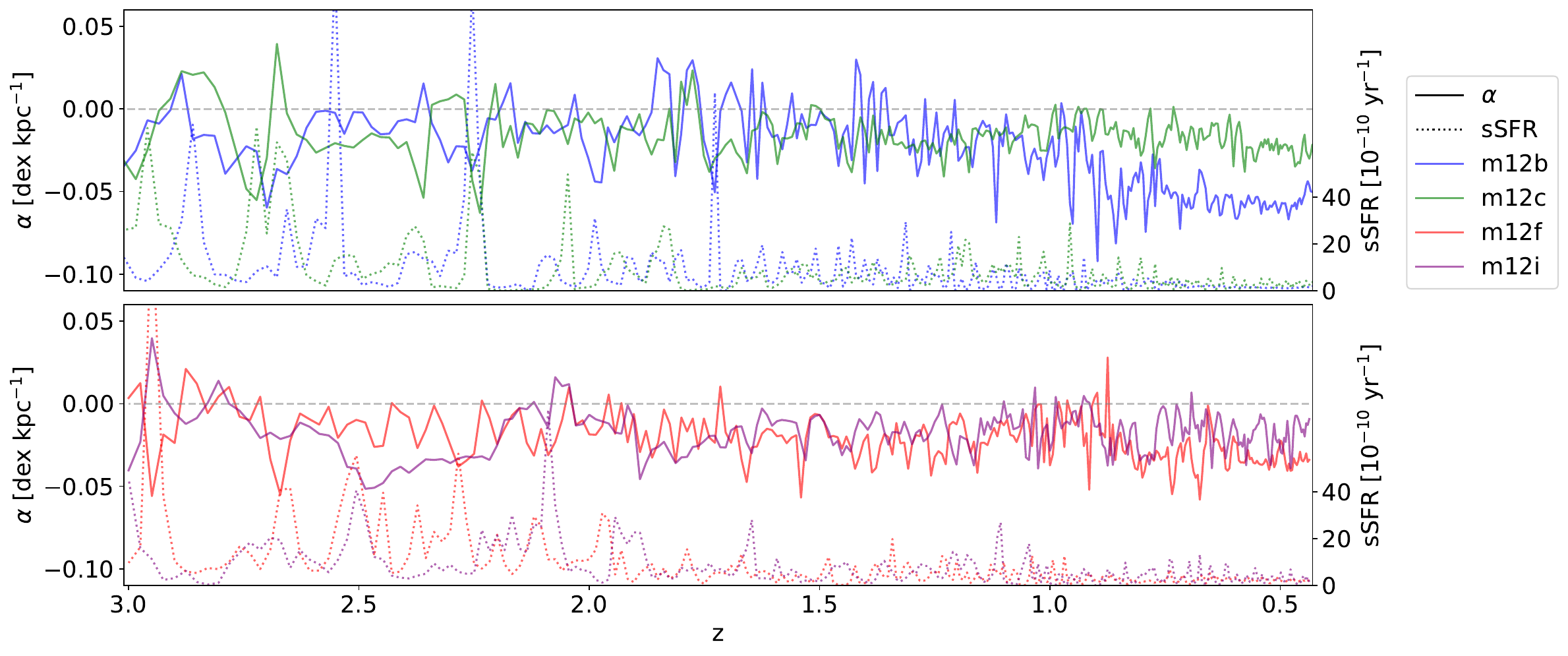}
 \includegraphics[width=0.8\linewidth]{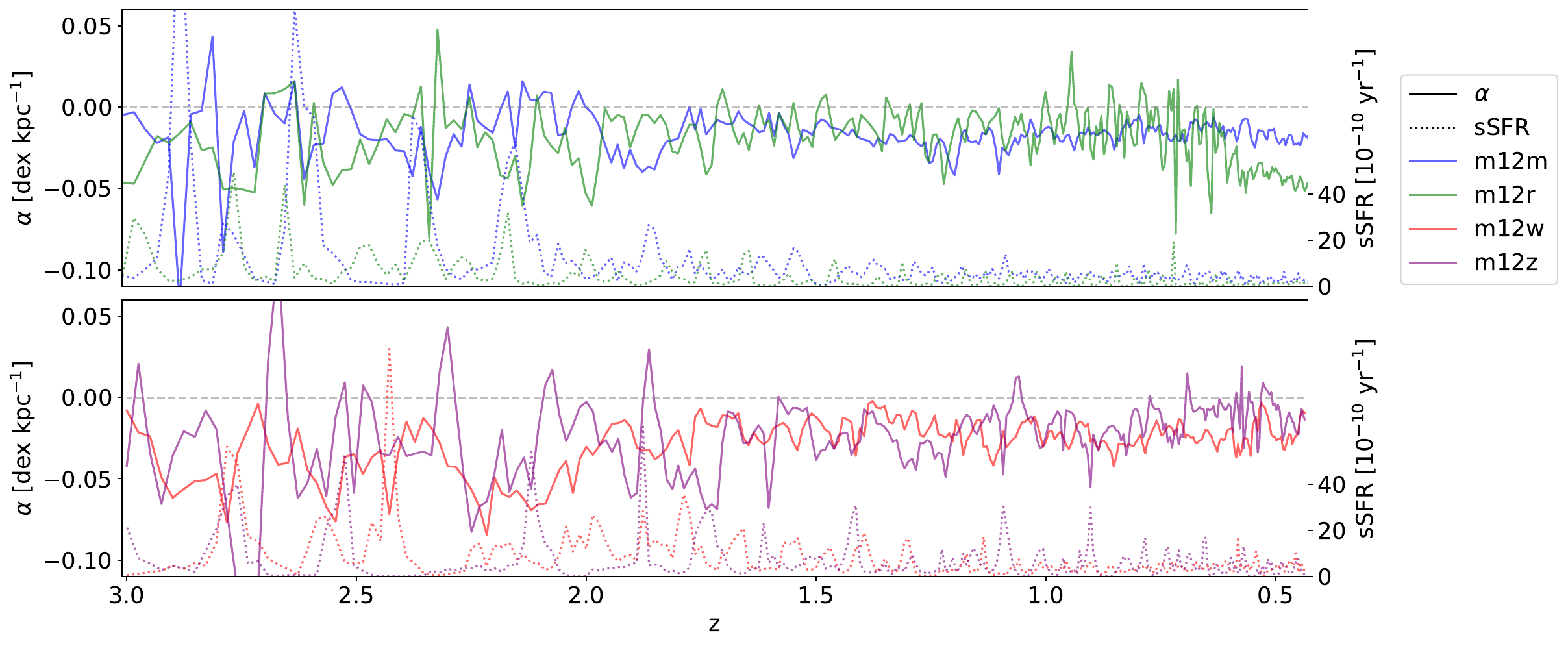}
 \caption{\emph{Evolution of metallicity gradient and sSFR in galaxies \texttt{\rm{m12b}},\texttt{\rm{m12c}}, \texttt{\rm{m12f}}, \texttt{\rm{m12i}}, \texttt{\rm{m12m}}, \texttt{\rm{m12r}}, \texttt{\rm{m12w}}, \texttt{\rm{m12z}}.}}
\label{fig:alphatozs}
\end{figure*}

Also, we have shown the evolution of gradient and sSFR for all 8 galaxies, as illustrated in Fig.~\ref{fig:alphatozs}, there are significant differences in the evolution of different galaxies.

\clearpage
\bibliography{sample631}{}
\bibliographystyle{aasjournal}

\clearpage

\begin{table*}[!htb]
    \label{tab:galadata}
    \caption{Galaxy properties, metallicity gradients and kinematics of the simulated sample.}
    \centering
    \setlength{\tabcolsep}{21pt} 
\begin{threeparttable}
\begin{tabular*}{0.95\linewidth}{@{}ccccccc@{}}
\hline
Name & $z$ & $R_{90}$ & $M_{\ast}$ & sSFR  & $\alpha$& $v_{\rm c}/\sigma$ \\
  &   & kpc & M$_\odot$ & 10$^{-11}$yr$^{-1}$& dex$\cdot$kpc$^{-1}$  &  \\
\hline
\texttt{m12b} & 0.5 & 5.12 & $ 5.31 \times 10^{10} $ & 9.983 & -0.057 $\pm$ 0.012 & 4.627 \\
\texttt{m12b} & 1.0 & 3.84 & $ 3.12 \times 10^{10} $ & 33.222 & -0.0166 $\pm$ 0.014 & - \\
\texttt{m12b} & 2.0 & 6.38 & $ 6.39 \times 10^{9} $ & 57.957 & -0.0297 $\pm$ 0.007 & 0.388 \\
\texttt{m12b} & 3.0 & 9.36 & $ 9.27 \times 10^{8} $ & 164.183 & -0.0345 $\pm$ 0.006 & 0.434 \\
\texttt{m12c} & 0.5 & 6.34 & $ 2.97 \times 10^{10} $ & 18.847 & -0.0213 $\pm$ 0.012 & 2.097 \\
\texttt{m12c} & 1.0 & 10.0 & $ 1.38 \times 10^{10} $ & 57.339 & -0.0032 $\pm$ 0.009 & 0.798 \\
\texttt{m12c} & 2.0 & 5.02 & $ 1.79 \times 10^{9} $ & 55.109 & -0.0022 $\pm$ 0.011 & 0.839 \\
\texttt{m12c} & 3.0 & 6.16 & $ 4.22 \times 10^{8} $ & 191.522 & -0.0314 $\pm$ 0.007 & 1.185 \\
\texttt{m12f} & 0.5 & 7.14 & $ 4.56 \times 10^{10} $ & 19.564 & -0.0224 $\pm$ 0.013 & 2.825 \\
\texttt{m12f} & 1.0 & 5.02 & $ 2.59 \times 10^{10} $ & 28.152 & -0.0374 $\pm$ 0.011 & 2.117 \\
\texttt{m12f} & 2.0 & 5.94 & $ 8.31 \times 10^{9} $ & 98.645 & -0.0186 $\pm$ 0.009 & 0.544 \\
\texttt{m12f} & 3.0 & 9.14 & $ 1.06 \times 10^{9} $ & 175.587 & 0.0035 $\pm$ 0.005 & 1.214 \\
\texttt{m12i} & 0.5 & 7.76 & $ 3.91 \times 10^{10} $ & 24.349 & -0.0293 $\pm$ 0.01 & 2.652 \\
\texttt{m12i} & 1.0 & 6.3 & $ 1.81 \times 10^{10} $ & 52.143 & -0.0285 $\pm$ 0.01 & - \\
\texttt{m12i} & 2.0 & 5.34 & $ 3.42 \times 10^{9} $ & 150.922 & -0.0067 $\pm$ 0.011 & 0.451 \\
\texttt{m12i} & 3.0 & 6.1 & $ 6.44 \times 10^{8} $ & 304.983 & -0.0403 $\pm$ 0.008 & 1.046 \\
\texttt{m12m} & 0.5 & 5.96 & $ 5.68 \times 10^{10} $ & 24.372 & -0.0216 $\pm$ 0.017 & 2.397 \\
\texttt{m12m} & 1.0 & 7.04 & $ 2.00 \times 10^{10} $ & 50.434 & -0.0118 $\pm$ 0.011 & 1.862 \\
\texttt{m12m} & 2.0 & 8.52 & $ 1.86 \times 10^{9} $ & 94.827 & 0.0003 $\pm$ 0.006 & 0.747 \\
\texttt{m12m} & 3.0 & 8.66 & $ 2.86 \times 10^{8} $ & 180.445 & -0.0048 $\pm$ 0.005 & 0.218 \\
\texttt{m12r} & 0.5 & 5.86 & $ 6.17 \times 10^{09} $ & 9.25 & -0.0426 $\pm$ 0.015 & 2.412 \\
\texttt{m12r} & 1.0 & 3.04 & $ 4.45 \times 10^{9} $ & 18.591 & -0.0035 $\pm$ 0.014 & 0.132 \\
\texttt{m12r} & 2.0 & 3.48 & $ 2.54 \times 10^{9} $ & 28.791 & -0.0523 $\pm$ 0.011 & 1.685 \\
\texttt{m12r} & 3.0 & 7.92 & $ 1.10 \times 10^{9} $ & 16.261 & -0.0462 $\pm$ 0.006 & 0.914 \\
\texttt{m12w} & 0.5 & 3.18 & $ 2.11 \times 10^{10} $ & 23.728 & -0.0263 $\pm$ 0.027 & 2.005 \\
\texttt{m12w} & 1.0 & 3.16 & $ 8.71 \times 10^{9} $ & 29.567 & -0.027 $\pm$ 0.015 & 0.992 \\
\texttt{m12w} & 2.0 & 9.42 & $ 1.66 \times 10^{9} $ & 84.956 & -0.0302 $\pm$ 0.006 & 1.377 \\
\texttt{m12w} & 3.0 & 6.88 & $ 3.42 \times 10^{8} $ & 121.815 & -0.0079 $\pm$ 0.009 & 0.353 \\
\texttt{m12z} & 0.5 & 6.0 & $ 6.84 \times 10^{9} $ & 25.892 & -0.0078 $\pm$ 0.024 & 0.866 \\
\texttt{m12z} & 1.0 & 8.16 & $ 2.46 \times 10^{9} $ & 51.272 & -0.0235 $\pm$ 0.008 & 0.364 \\
\texttt{m12z} & 2.0 & 8.08 & $ 4.27 \times 10^{8} $ & 67.919 & -0.0027 $\pm$ 0.008 & 1.065 \\
\texttt{m12z} & 3.0 & 4.58 & $ 2.05 \times 10^{8} $ & 113.291 & -0.0419 $\pm$ 0.023 & 0.031 \\
\hline
\end{tabular*}
\begin{tablenotes}
\item \hspace{-13pt}Galaxy properties studied in this paper (units are physical):
\item Name: simulation designation.
\item $z$: redshift where the properties here are measured.
\item $R_{90}$: the radius that contains 90 percent of the SFR (averaged over 200 Myr) density.
\item $M_{\ast}$: stellar mass within the central 10 kpc of the galaxy at the given redshift.
\item sSFR: specific star formation rate within the central 10 kpc of the galaxy (averaged over 200 Myr).
\item $\alpha$: gas-phase metallicity gradient measured over 0.25-1$R_{90}$.
\item $v_{\rm c}/\sigma$: degree of rotational support for the ISM gas.
\end{tablenotes}
\end{threeparttable}
\end{table*}

\end{document}